\documentclass[twocolumn,showpacs,preprintnumbers,amsmath,amssymb,nofootinbib,aps,prc,10pt]{revtex4-1}
\usepackage{graphicx}
\usepackage{dcolumn}
\usepackage{bm}

\usepackage{pstricks}
\usepackage{psboxit}
\usepackage{color}
\usepackage{hyperref}

\makeatletter
\def\HyPsd@CatcodeWarning#1{}
\makeatother

\begin{document}
\newcommand{\D}{{\rm d}}

\title{Phase space structure and dynamics within the time-dependent Hartree-Fock approach} 
\author{N. Loebl$^1$, A. S. Umar$^2$, J.A. Maruhn$^1$, P.-G. Reinhard$^3$, P.D. Stevenson$^4$, and V. E., Oberacker$^2$}
 
\affiliation{
$^1$Institut fuer Theoretische Physik, 
Universitaet Frankfurt, D-60438 Frankfurt, Germany 
} 
\affiliation{
$^2$Department of Physics and Astronomy, Vanderbilt University,
Nashville, Tennessee 37235, USA
}
\affiliation{
$^3$Institut fuer Theoretische Physik II,
Universitaet Erlangen-Nuernberg, D-91058 Erlangen, Germany
}
\affiliation{
$^4$Department of Physics, University of Surrey, Guildford, GU2 7HX, UK
}

\date{\today}

\begin{abstract}
We study the equilibration and relaxation processes within the 
time-dependent Hartree-Fock approach using the Wigner distribution function.
On the technical side we present a geometrically unrestricted
framework which allows us to calculate the full six-dimensional Wigner
distribution function. With the removal of geometrical constraints, we
are now able to extend our previous phase-space analysis of heavy-ion
collisions in the reaction plane to unrestricted mean-field simulations of nuclear matter on a
three-dimensional Cartesian lattice. From the physical point of view we provide a quantitative analysis
on the stopping power in TDHF. This is linked to the effect of transparency. For the medium-heavy
$^{40}$Ca+$^{40}$Ca system we examine the impact of different
parametrizations of the Skyrme force, energy-dependence, and the significance of extra time-odd terms
in the Skyrme functional. For the first time, transparency in TDHF is observed for
a heavy system, $^{24}$Mg+$^{208}$Pb.
\end{abstract}

\pacs{21.60.-n,21.60.Jz}

\maketitle

\section{Introduction}

Time-dependent Hartree-Fock (TDHF) theory provides a fully self-consistent mean-field approach to nuclear dynamics.
First employed in the late 1970's~\cite{Bonche,Svenne,Negele,Davies} the
applicability of TDHF was constrained by the limited
computational power. Therefore, early applications 
treated the problem in only one spatial dimension, utilizing a very
simplified parametrization of the nuclear interaction.
Due to the increase in computational power, state-of-the-art
TDHF calculations are now feasible in three-dimensional
coordinate space, without any symmetry restrictions and using the full Skyrme 
interaction~\cite{Kim,Simenel,Nakatsukasa,Umar05a,Maruhn1,Guo08a}.

In this work the Wigner distribution function~\cite{Wigner} is
calculated as an analysis tool to probe the phase space behavior in
TDHF evolution of nuclear dynamics. 
In comparison to previous work~\cite{Loebl}, where the Wigner 
analysis was performed in one and two dimensions,
we are now able to carry out both the TDHF simulation
and the phase-space analysis in three dimensions. 
Transformation from 
coordinate-space representation to phase-space representation, i.e. calculating the Wigner
distribution from the density matrix, still remains a computationally challenging problem.
Here, we
present a fully three-dimensional analysis which allows the study of
relaxation processes simultaneously in all directions in $k$-space.
An early one-dimensional study of the Wigner function for
TDHF can be found in~\cite{Maruhn3}.  

The paper is outlined as follows: In Sec.~\ref{sec:wigner} we
introduce the Wigner distribution function and discuss the
numerical framework used in this work. Some benchmark results are
presented to give an idea about the computational
resources necessary to compute the Wigner function. We then
introduce the principal observables summarizing the local or global momentum-space properties
of the Wigner function. First the quadrupole operator in momentum space 
which gives rise to the usual deformation parameters $\beta$ and $\gamma$ to
probe relaxation procecces in dynamical calculations.
In addition, we define an estimate for the occupied phase-space volume to
obtain a relation between the fragment separation in momentum and
coordinate space. 
 
Sec.~\ref{sec:results} illustrates the geometrical structure of the
Wigner distribution by means of momentum cuts for a few static
nuclei. This is followed
by a detailed discussion of the central $^{40}$Ca+$^{40}$Ca collision, paying
particular attention to the effect of transparency. We discuss
the impact of different Skyrme parametrizations on the 
relaxation behavior, as well as the dependence on the center-of-mass energy
for a fixed Skyrme interaction. We also examine the influence of extra
time-odd terms in the Skyrme functional. We complete this issue by 
looking at the $^{24}$Mg+$^{208}$Pb reaction, where transparency can be studied
for the case of a heavy system. 

\section{Outline of formalism}
\label{sec:wigner}

\subsection{Solution of the TDHF equations}

The TDHF equations are solved on a three-dimensional Cartesian lattice
with a typical mesh spacing of $1$\:fm. The initial
setup of a dynamic calculation needs a static Hartree-Fock run, whereby the
stationary ground states of the two fragments are computed with the
damped-gradient iteration algorithm~\cite{Blum,Reinhard}. 
The TDHF runs are initialized with energies above the Coulomb barrier at some
large but finite separation. The two ions are boosted with velocities obtained by assuming
that the two nuclei arrive at this initial separation on a Coulomb trajectory.
The time propagation is managed by
utilizing a Taylor-series expansion of the time-evolution operator~\cite{Flocard}
up to sixth order with a time step of $t=0.2$\:fm/c.
The spatial derivatives are calculated using the fast Fourier transforms (FFT).

\subsection{Computing the Wigner function}

The Wigner distribution function is obtained by a partial Fourier
transform of the density matrix
$\rho(\mathbf{r}\!-\!\frac{\mathbf{s}}{2},    
\mathbf{r}\!+\!\frac{\mathbf{s}}{2},t)$, with respect to the 
relative coordinate $\mathbf{s}=\mathbf{r}-\mathbf{r}'$

\begin{eqnarray}
 f^{(3)}_\mathrm{W}(\mathbf{r},\mathbf{k},t)
 &=&
  \int\frac{{\rm d}^3 s}{(2\pi)^3}\:
  e^{-i\mathbf{k}\cdot\mathbf{s}} 
  \rho(\mathbf{r}\!-\!\frac{\mathbf{s}}{2},
       \mathbf{r}\!+\!\frac{\mathbf{s}}{2},t)
  \;,
\\
  \rho(\mathbf{r},\mathbf{r}',t)
  &=&
  \sum_{l}\Psi^\dagger_{l}(\mathbf{r},t)\Psi_{l}(\mathbf{r}',t)
\:.
\end{eqnarray}
Because $f_\mathrm{W}$ is not
positive definite, it is misleading to consider the Wigner function as a phase-space probability
distribution. We will refer to the appearance of negative
values for $f_\mathrm{W}$ in Sec.~\ref{sec:results}.

Evaluating the Wigner function in six-dimensional phase space is still a computational  challenge and only possible 
employing full Open MP parallelization and extensive use of FFT's. The determing factor is the grid size, which results in
\begin{equation}
N_{x}^{2}\log{(N_x)}\star N_{y}^{2}\log{(N_y)}\star
N_{z}^{2}\log{(N_z) }
\end{equation}
steps to provide the Wigner transform in full
space, where $N_x,N_y,N_z$ are the grid points in each direction. 

\begin{figure}[hbtp]
\centering
  \newsavebox\IBox
\savebox\IBox{   \includegraphics*[width=8.1cm]{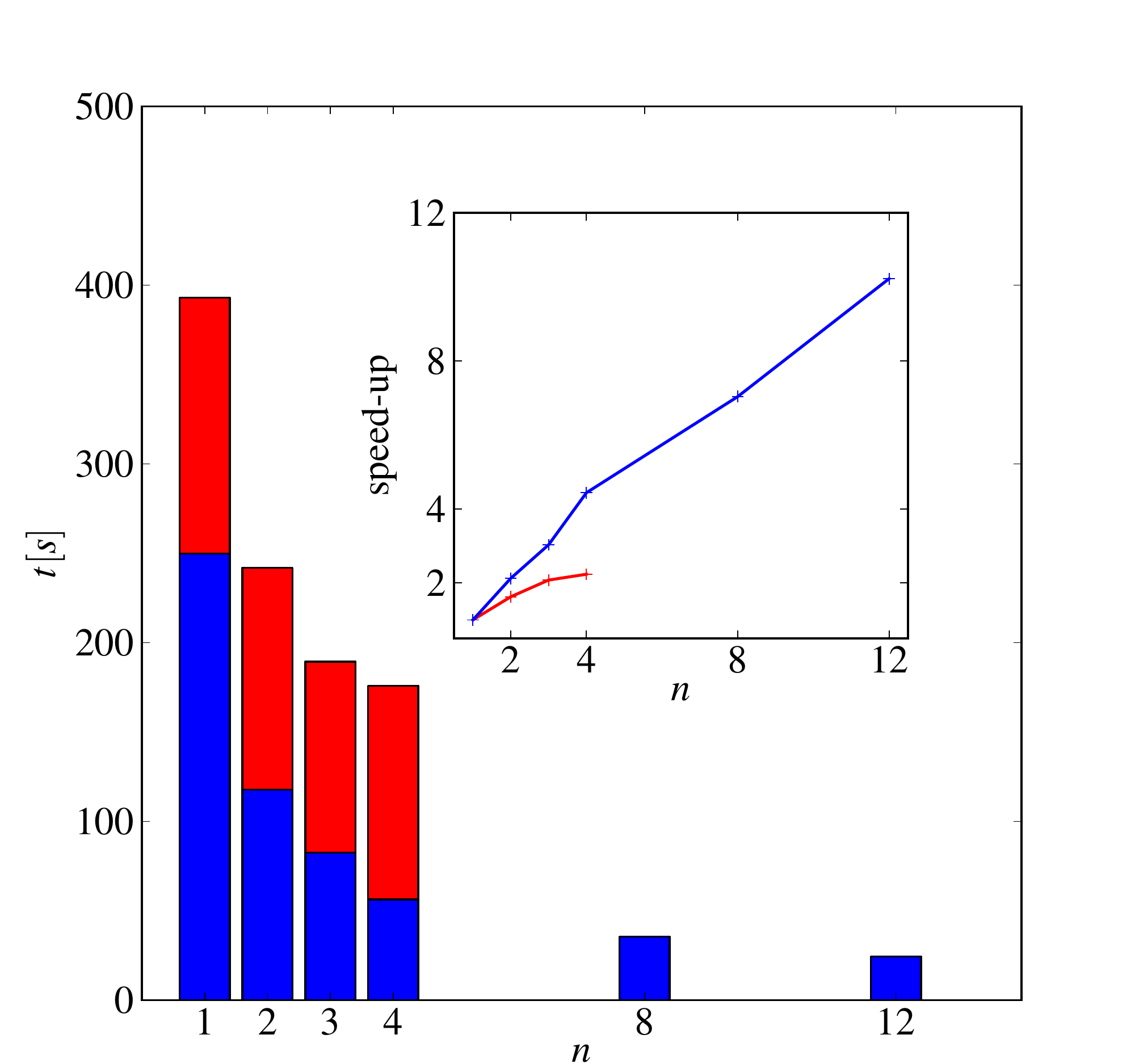}}
 \begin{pspicture}(\wd\IBox,\ht\IBox)
    \rput[lb](-0.5,0.5){\usebox\IBox}
    \rput[lb](0.8,6.7){(a)}  
    \rput[lb](3.2,6.){(b)}  
  \end{pspicture}
\caption{\label{fig:O16_runtime} (color online)
Wall clock time $t$ in seconds (a) and speed-up (b) by evaluating the Wigner function at one
single time step on a $24^2\times36$ grid for $^{16}O+$$^{16}O$ as a
function of the number of used processors $n$. CPU's used: (CPU):  
Intel(R) Core(TM)2 Quad CPU 2.66GHz,  Cores:  4 (red), Intel(R)
Xeon(R) CPU X5680 3.33GHz, Cores: 12 (blue).}
\end{figure}
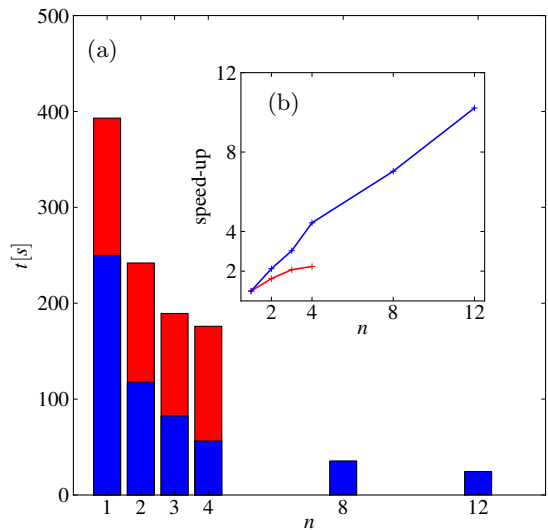

Figure~\ref{fig:O16_runtime} shows the time taken to evaluate
the Wigner function for one single time-step on a $24^2\times36$ grid
for a $^{16}$O+$^{16}$O collision. Benchmarks are shown for two
different CPU's. 
Storing the Wigner function reduced to the reaction
plane, i.e. $f^{(3)}_\mathrm{W}(x,y=0,z,\mathbf{k})$ at one time step will consume
$\sim 140$\;Mb of disk space for the presented case in Fig.~\ref{fig:O16_runtime}. Going to larger grid sizes, 
needed for heavier systems,
and/or storing the full three-dimensional Wigner function will clearly result in entering
the Gb regime. 

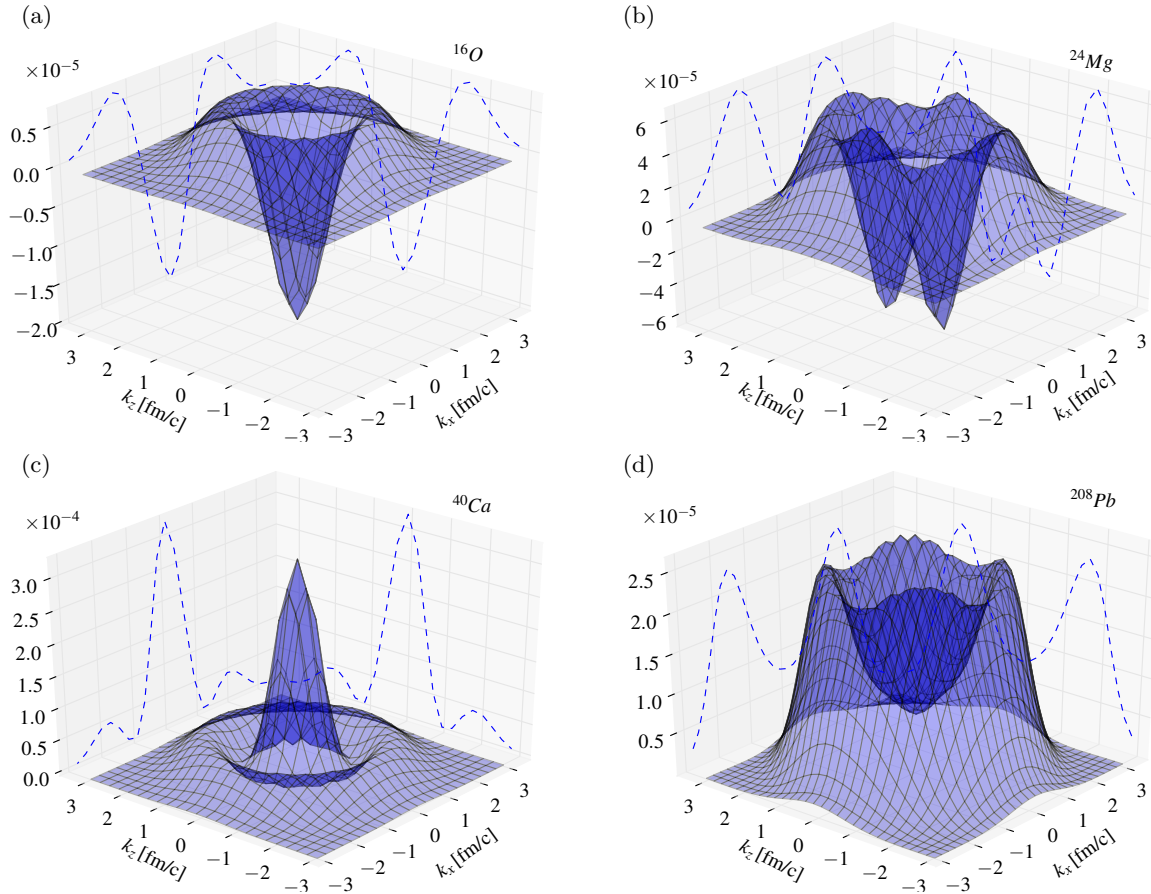
\begin{figure*}[hbtp]
 \centering

  \savebox\IBox{\includegraphics*[width=8.1cm]{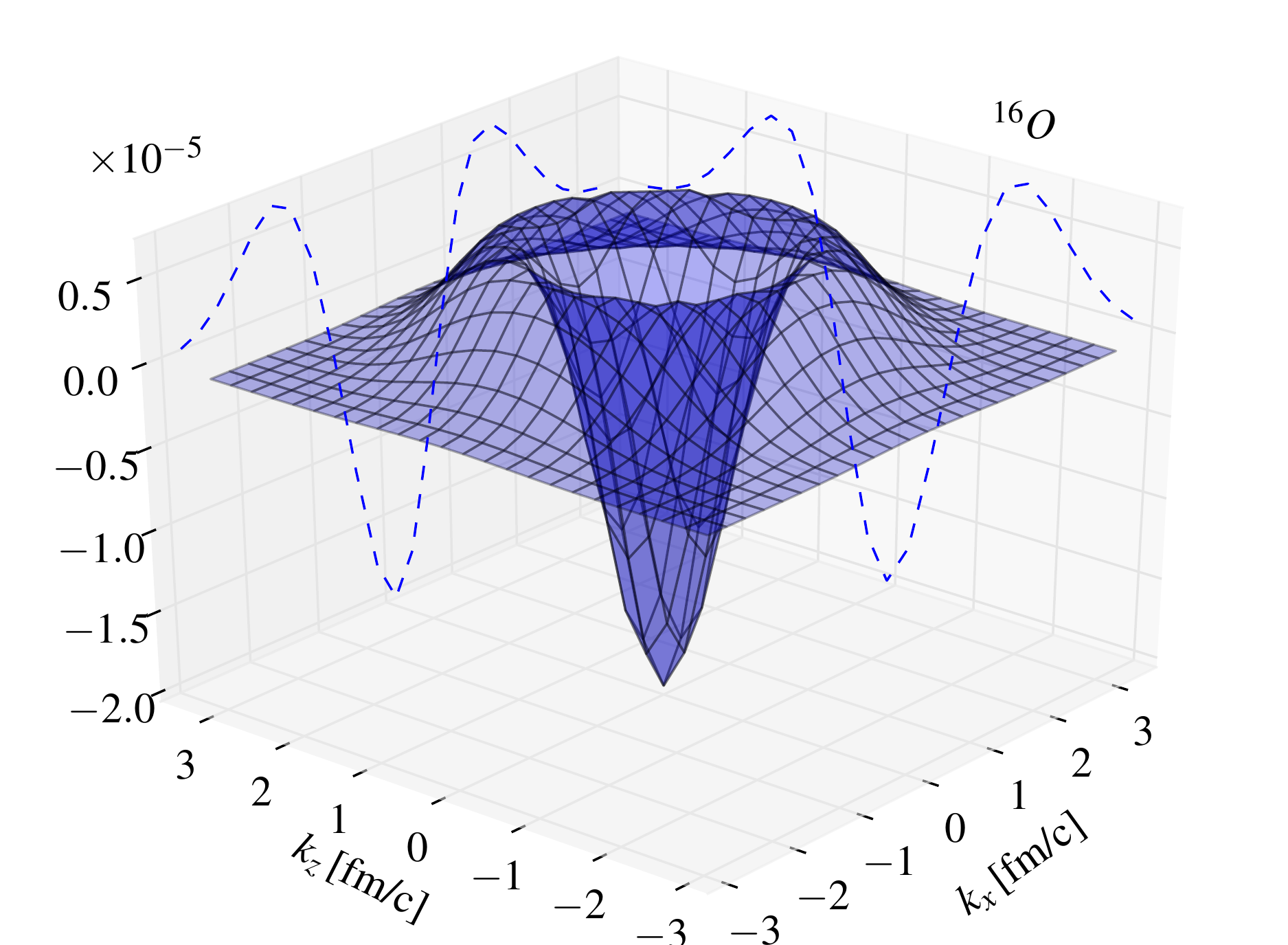}
		\includegraphics*[width=8.1cm]{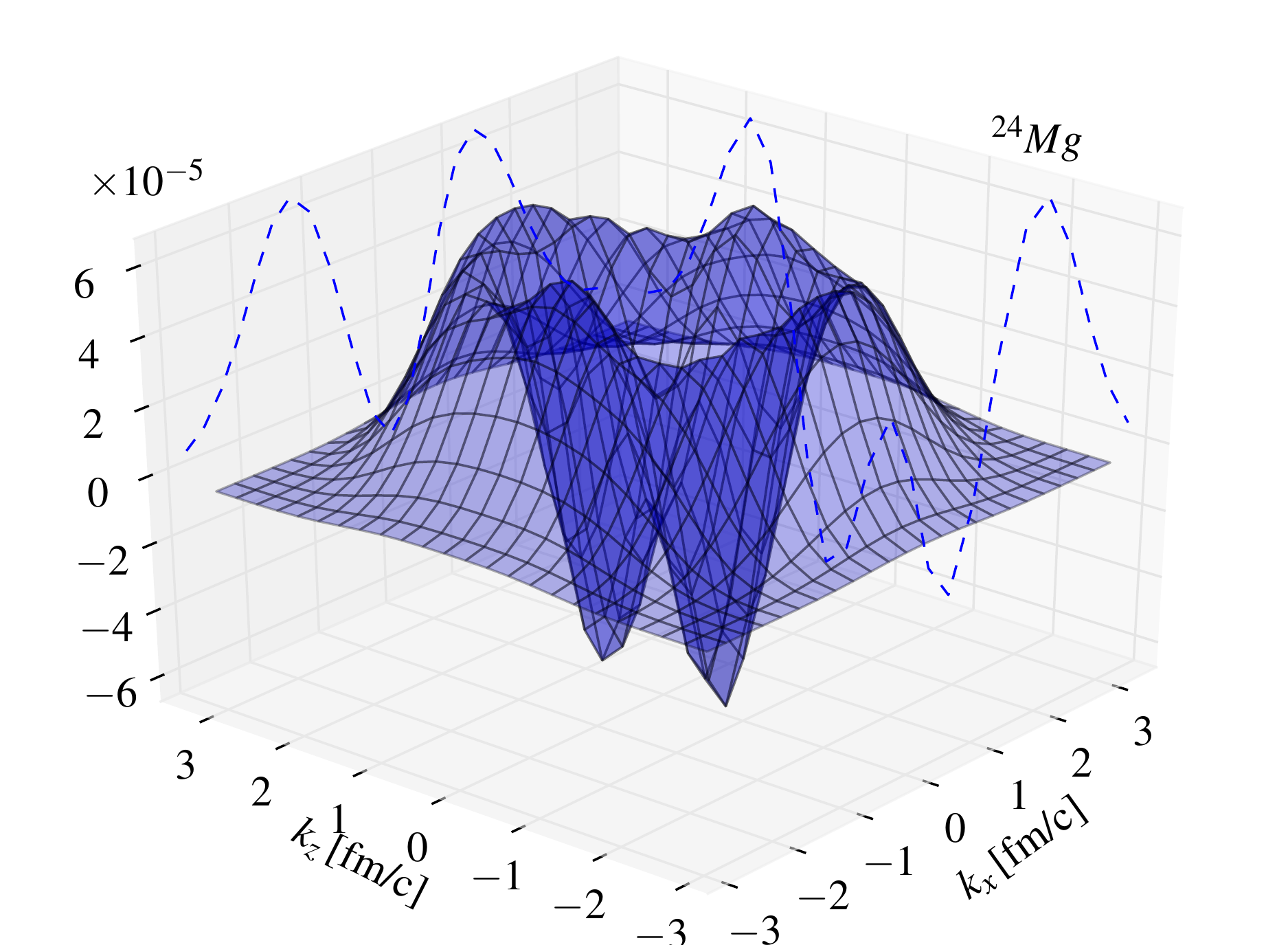}}
 \begin{pspicture}(\wd\IBox,\ht\IBox)
    \rput[lb](-0.5,0.5){\usebox\IBox}
    \rput[lb](0.,6.){(a)}  
    \rput[lb](8.,6.){(b)} 
  \end{pspicture}
  \savebox\IBox{\includegraphics*[width=8.1cm]{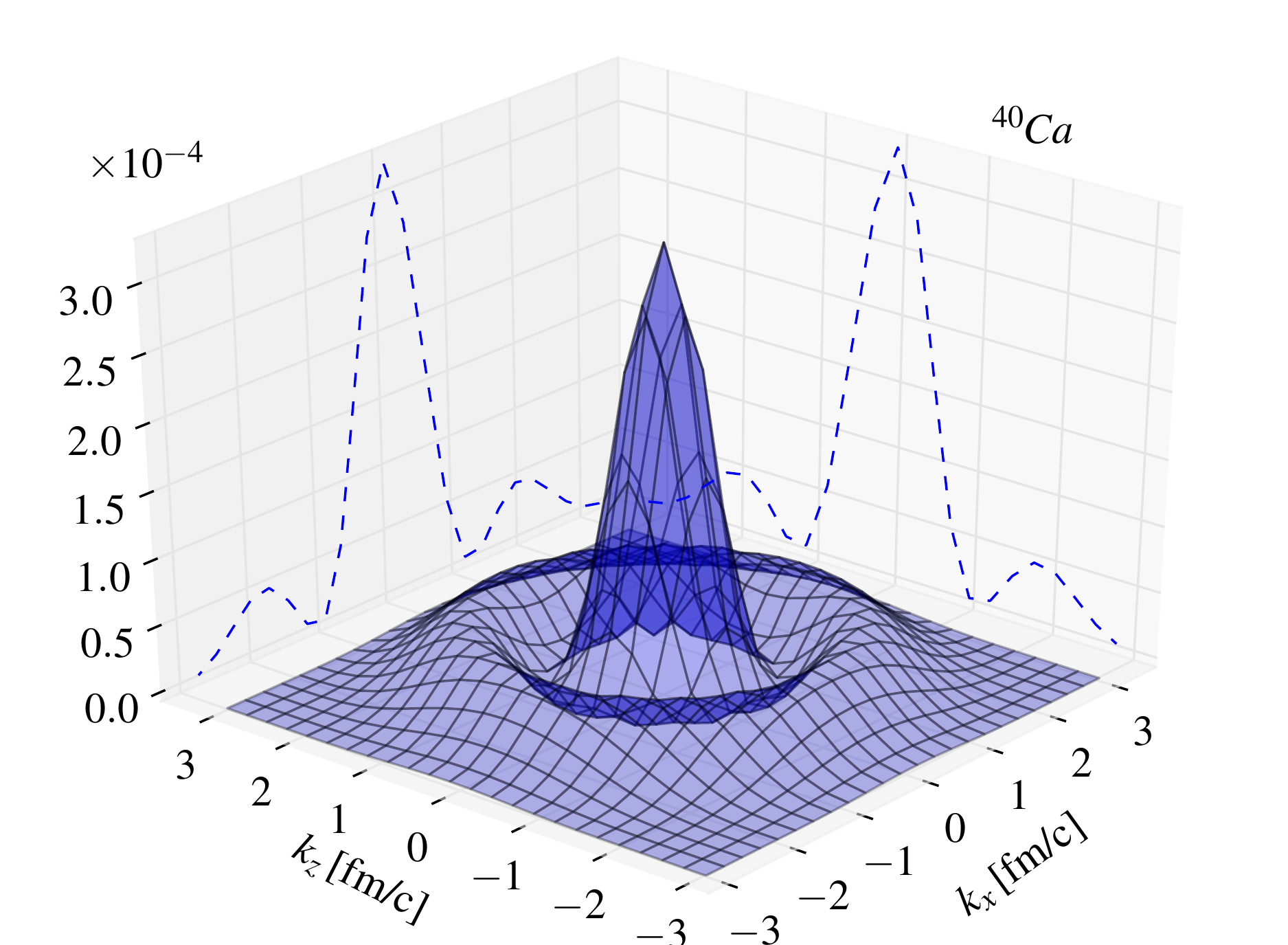}
		\includegraphics*[width=8.1cm]{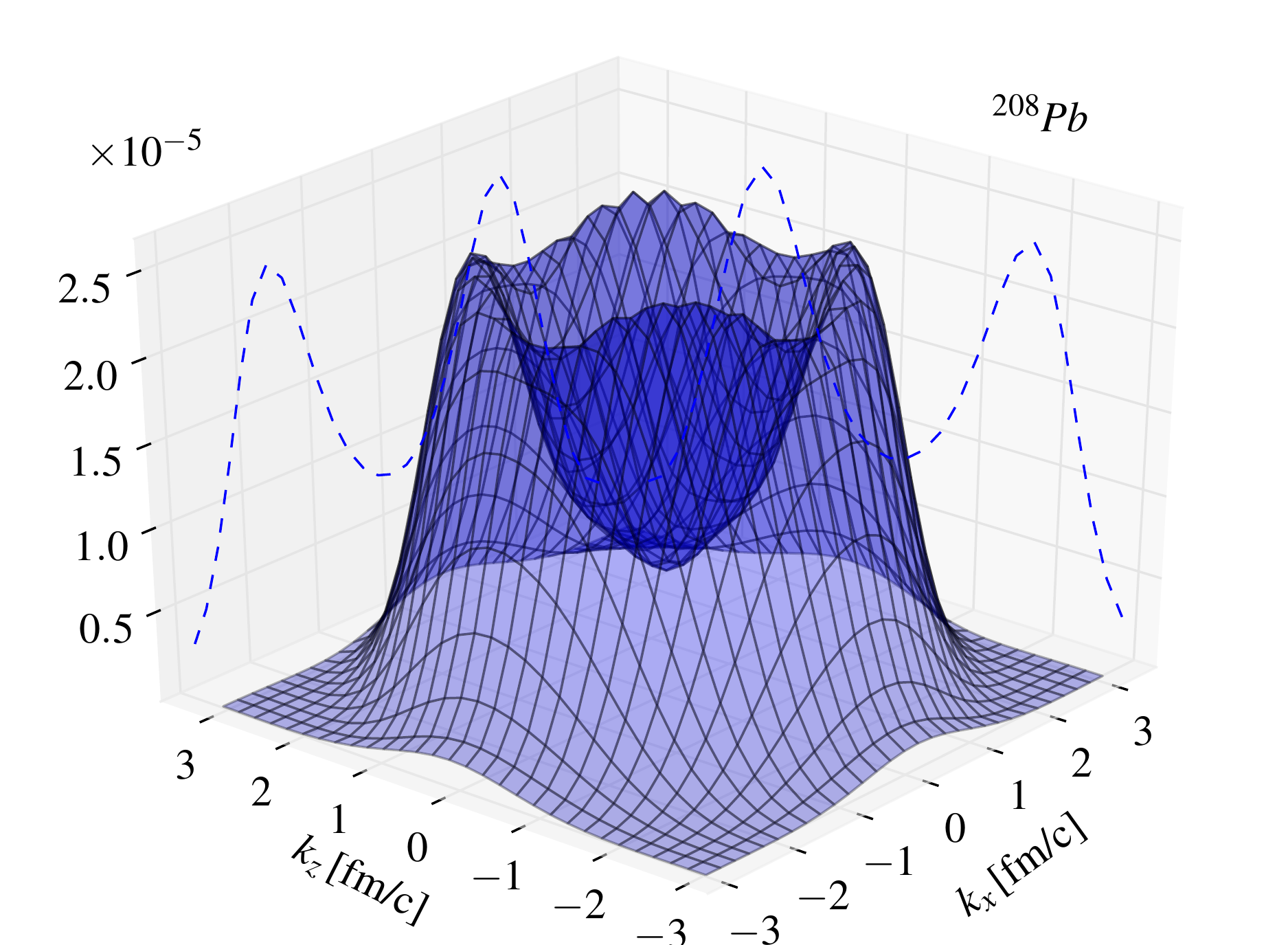}}
 \begin{pspicture}(\wd\IBox,\ht\IBox)
    \rput[lb](-0.5,0.5){\usebox\IBox}
    \rput[lb](0.,6.){(c)}  
    \rput[lb](8.,6.){(d)} 
  \end{pspicture}
\caption{\label{fig:Static} (color online)
Two-dimensional $k_x$-$k_z$ subspace from the full Wigner distribution
function $f^{(3)}(\mathbf{r},\mathbf{k})$ in the spatial center of
the static nuclei $^{16}O$ (a), $^{24}Mg$ (b), $^{40}Ca$ (c), and
$^{208}Pb$ (d).
One-dimensional cuts along the $k_x$/$k_z$-axis are projected at the
walls. The Skyrme interaction SLy6 was used for all cases.}
\end{figure*}

\subsection{Observables}
\label{sec:observ}

In this section we discuss some of the observables used in our analysis.
In order to avoid any misunderstandings we will label all observables evaluated in momentum space with
a subscript $k$, and all observables in coordinate space with a
subscript $r$.

\subsubsection{Quadrupole in momentum space}

As an observable to probe relaxation in phase-space quantitatively, we
evaluate the quadrupole operator in momentum space. The local
deviation of the momentum distribution from a spherical shape is a
direct measure for equilibration. The local quadrupole tensor in
$k$-space is given by
\begin{equation}
Q^{ij}_k(\mathbf{r},t)=\int \D^3 k
\left[ 3\langle k_i(\mathbf{r},t) \rangle \langle k_j(\mathbf{r},t)
\rangle - \langle \mathbf{k}^2(\mathbf{r},t) \rangle
\delta_{ij}\right]\:,
\end{equation}
using the $m$-th moment from the local momentum
distribution 
\begin{equation}
  \langle\mathbf{k}^{(m)}(\mathbf{r},t)\rangle
  = 
  \frac{\int {\rm d}^3 k \:(\mathbf{k}-\langle\mathbf{k}(\mathbf{r},
          t)\rangle)^{m}f^{(3)}_\mathrm{W}(\mathbf{r},\mathbf{k},t)}
       {\int {\rm d}^3
k\:f^{(3)}_\mathrm{W}(\mathbf{r},\mathbf{k},t)}
  \:,
\end{equation}
with  $\langle\mathbf{k}(\mathbf{r}, t)\rangle$ denoting the average
local flow
\begin{equation}
  \langle\mathbf{k}(\mathbf{r},t)\rangle
  = 
  \frac{\int {\rm d}^3
k\:\mathbf{k}\:f^{(3)}_\mathrm{W}(\mathbf{r},\mathbf{k},t)}
       {\int {\rm d}^3
k\:f^{(3)}_\mathrm{W}(\mathbf{r},\mathbf{k},t)}\:.
\end{equation}
The spherical quadrupole moments $Q_k^{20}(\mathbf{r},t)$ and
$Q_k^{22}(\mathbf{r},t)$ are computed by diagonalization of
$Q_k^{ij}(\mathbf{r},t)$
\begin{eqnarray}
\label{eq:diag}
Q_k^{20}(\mathbf{r},t)&=&\sqrt{\frac{5}{16\pi}}\lambda_3  \\
Q_k^{22}(\mathbf{r},t)&=&\sqrt{\frac{5}{96\pi}}(\lambda_2 - \lambda_1)
\end{eqnarray}
with $\lambda_3 > \lambda_2 > \lambda_1$ labeling the eigenvalues of
$Q_k^{ij}(\mathbf{r},t)$. Switching to polar notation the
observables
\begin{eqnarray}
\beta_k(\mathbf{r},t)&=&
\sqrt{\beta_{20}^2(\mathbf{r},t)+2\beta_{22}^2(\mathbf{r},t)} \\
\gamma_k(\mathbf{r},t)&=&
|\arctan{\frac{\sqrt{2}\beta_{22}(\mathbf{r},t)}{\beta_{20}(\mathbf{r}
,t)}}\frac{ 180^{\circ}
}{\pi}|\: ,
\end{eqnarray}
are obtained via the dimensionless quantities
\begin{eqnarray}
\beta_k^{20}(\mathbf{r},t)&=&\frac{4\pi Q_{20}}{5 r_k^2
\rho(\mathbf{r},t)} \\
\beta_k^{22}(\mathbf{r},t)&=&\frac{4\pi Q_{22}}{5 r_k^2
\rho(\mathbf{r},t)}\:,
\end{eqnarray}
where
\begin{equation}
r_{k}(\mathbf{r},t)=\sqrt{\langle\mathbf{k}(\mathbf{r},
t)\rangle^2/\rho(\mathbf{r}, t) }\: ,
\end{equation}
accounts for the local rms-radius in $k$-space. The norm is defined such that
\begin{equation}
\rho(\mathbf{r},t)=\int \D^3
k\:f^{(3)}_\mathrm{W}(\mathbf{r},\mathbf{k},t)\:.
\end{equation}
In the presented formalism it is straightforward to define global
observables. The global quadrupole tensor is calculated by
spatial integration
\begin{equation}
Q_k^{ij}(t)=\int \D^3 r\: 
\rho(\mathbf{r},t)Q^{ij}_k(\mathbf{r},t)\:.
\end{equation}
Applying the same diagonalization as in the local case (\ref{eq:diag})
we end up with a global definition for $\beta_k^{20}(t)$ and
$\beta_k^{22}(t)$. For the following results we will mainly use the
global definition since it is more compact and allows the
simultaneous visualization of multiple time-dependent observables. We
will however show some local results in Sec. \ref{sec:mgpb}.

\subsubsection{Quadrupole in coordinate space}

To illustrate the global development of a reaction, we will also use
the expectation value $Q^{20}_r\equiv\langle \hat{Q}^{20}_r\rangle$ of
the quadrupole operator in coordinate space.

\subsubsection{Occupied phase space volume}

To give a rough measure for the phase-space volume occupied by
the fragments during a heavy-ion collision
we assume a spherical shape of the local momentum distribution. Adding
up the $k$-spheres 
\begin{equation}
V_k(\mathbf{r,t})=\frac{4\pi}{3}\langle\mathbf{k}^2(\mathbf{r},
t)\rangle^{3/2}\: ,
\end{equation}
leads to the total occupied phase-space volume
\begin{equation}
V_k(t)=\int {\rm d}^3 r \: V_k(\mathbf{r,t})\:.
\end{equation}

\section{Results and discussion}
\label{sec:results}

It is the aim of this work to provide a quantitative analysis of the 
magnitude of relaxation processes occurring in TDHF. Therefore we will
vary a single reaction parameter, while all the other parameters are
fixed. The $^{40}$Ca+$^{40}$Ca-system provides a suitable test
case. The occurrence of transparency will also be examined
for the heavy $^{24}$Mg+$^{208}$Pb system. However, we first start
with a brief discussion concerning the geometrical structure of the
full Wigner distribution for static nuclei.

\subsection{Static nuclei}
\label{sec:static}

In this section we present slices through the full six-dimensional
Wigner distribution for some static nuclei. For this purpose the
Wigner distribution is plotted at the spatial center, and for fixed
momentum in the $y$-direction, i.e. $k_y=0$.

Starting with $^{16}$O, which corresponds to subplot (a) in  Fig. \ref{fig:Static}, the Wigner function reveals strong negative values at the center. The hole in the momentum distribution, which indicates the influence of distinct shell effects, was already visible in the lower-dimensional analysis \cite{Loebl}, however in that case it corresponded to small positive values. 

Considering heavier nuclei, we observe the appearance of two
holes for case (b), reflecting the deformation of
$^{24}$Mg in coordinate space. In contrast we find a pronounced peak
for $^{40}$Ca, case (c). The momentum distribution for
$^{208}$Pb in subplot (d) reveals again a central hole which this
time is positive. This agrees with the reduction in the
spatial density observed for $^{208}$Pb.

Although the geometrical structure differs from nucleus to nucleus, it
is important to note that for cold ground-state nuclei the
momentum distribution differs considerably from a Fermi distribution,
even for a heavy nucleus.

\begin{figure}[hbtp]
 \centering
  \savebox\IBox{\includegraphics[width=8.1cm]{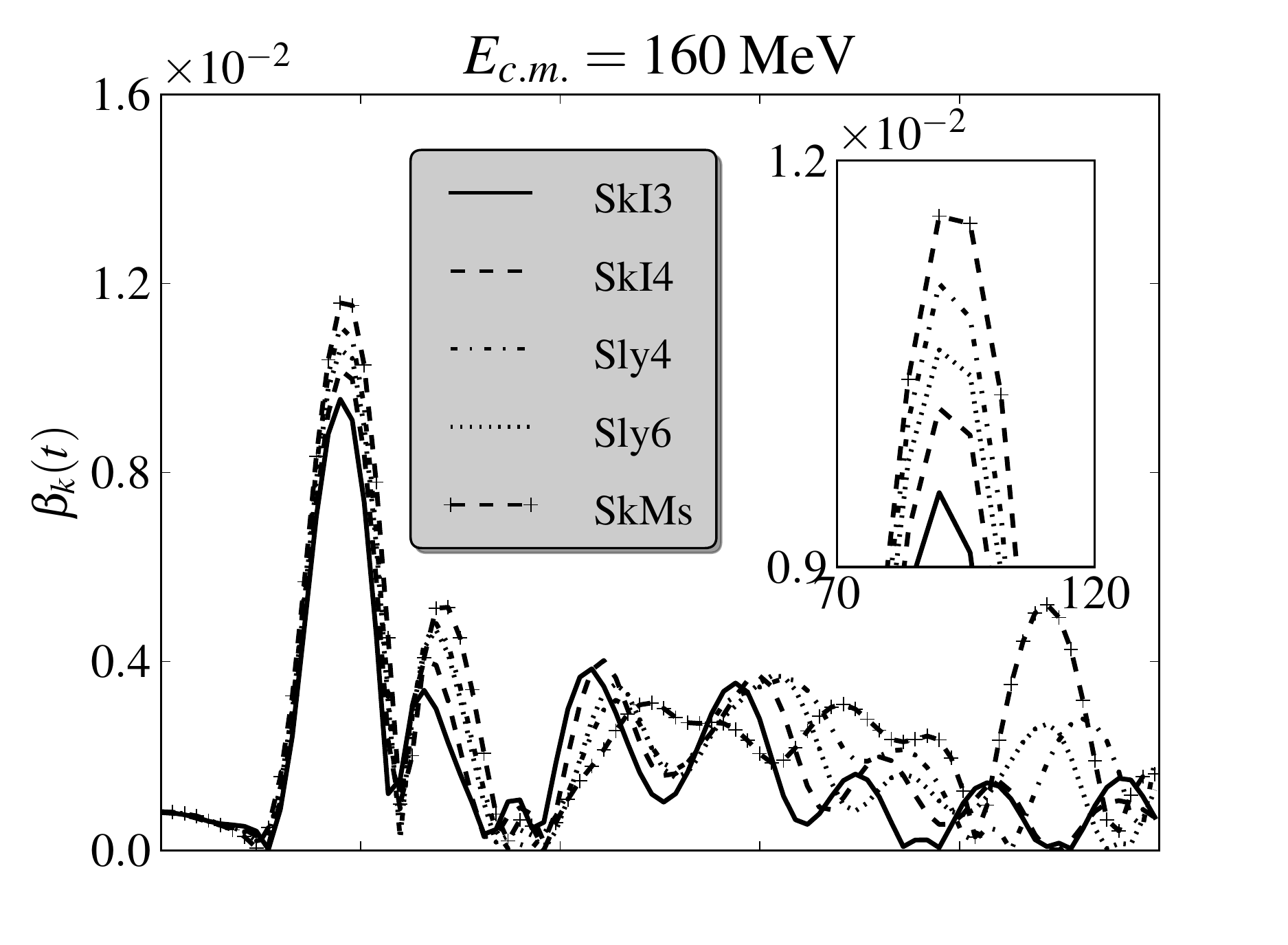}}
 \begin{pspicture}(\wd\IBox,\ht\IBox)
    \rput[lb](-0.5,0.5){\usebox\IBox}
    \rput[lb](-0.3,6.){(a)}  
    \rput[lb](4.8,5.){(e)}  
  \end{pspicture}
  \savebox\IBox{\includegraphics[width=8.1cm]{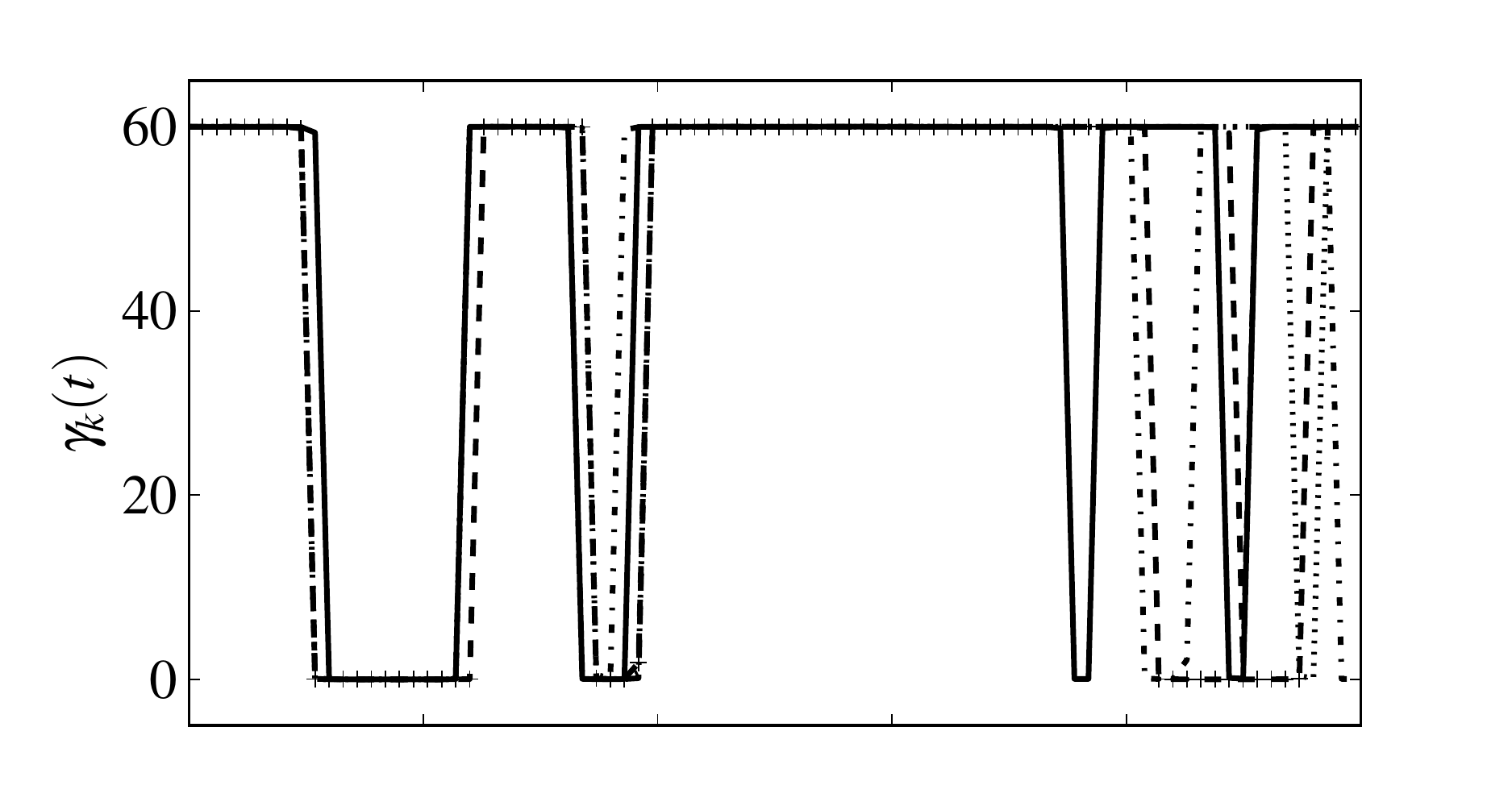}}
 \begin{pspicture}(\wd\IBox,\ht\IBox)
    \rput[lb](-0.5,0.5){\usebox\IBox}
    \rput[lb](-0.3,4.5){(b)}  
  \end{pspicture}
  \savebox\IBox{\includegraphics[width=8.1cm]{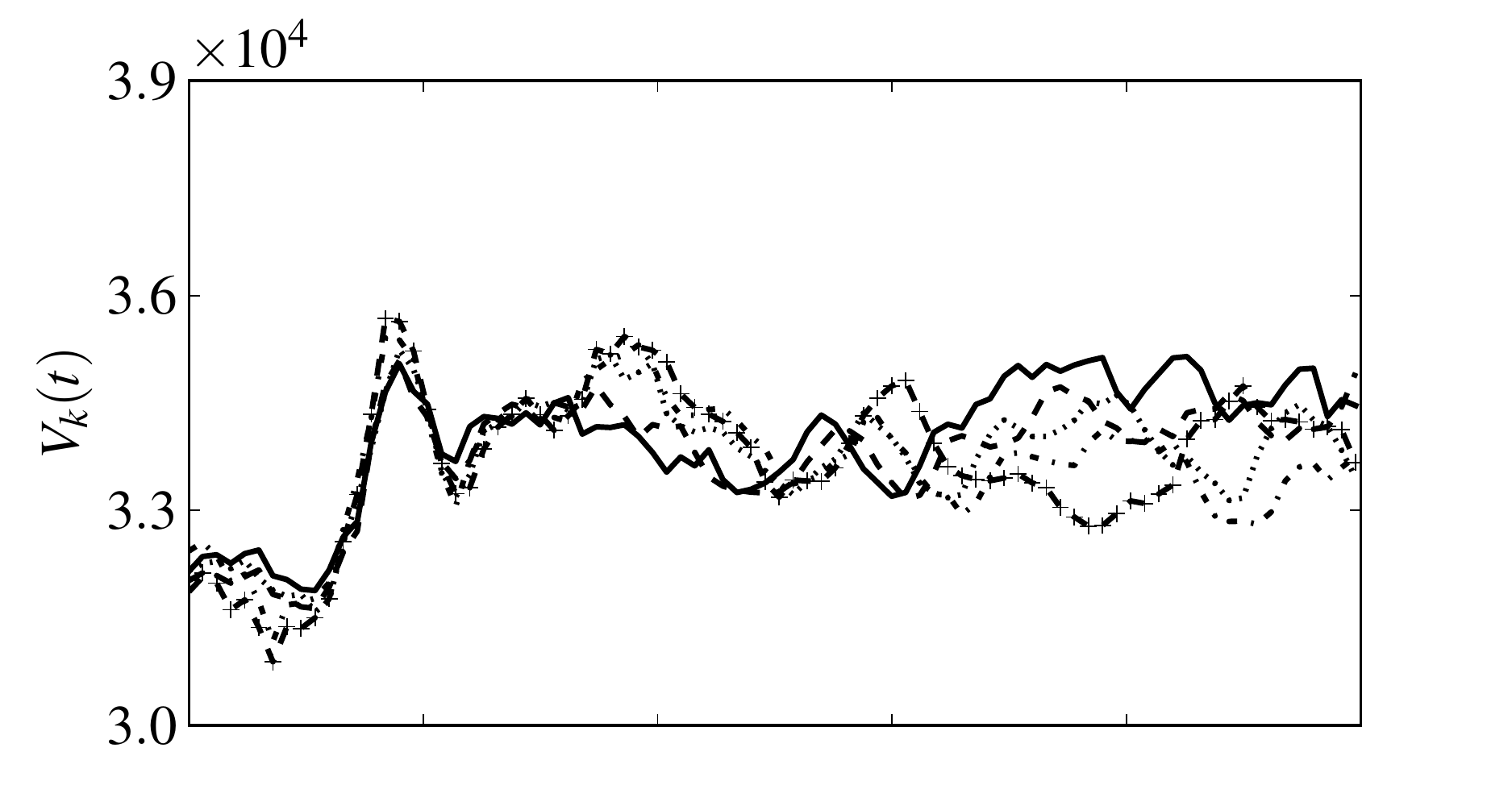}}
 \begin{pspicture}(\wd\IBox,\ht\IBox)
    \rput[lb](-0.5,0.5){\usebox\IBox}
    \rput[lb](-0.3,4.5){(c)}  
  \end{pspicture}
  \savebox\IBox{\includegraphics[width=8.1cm]{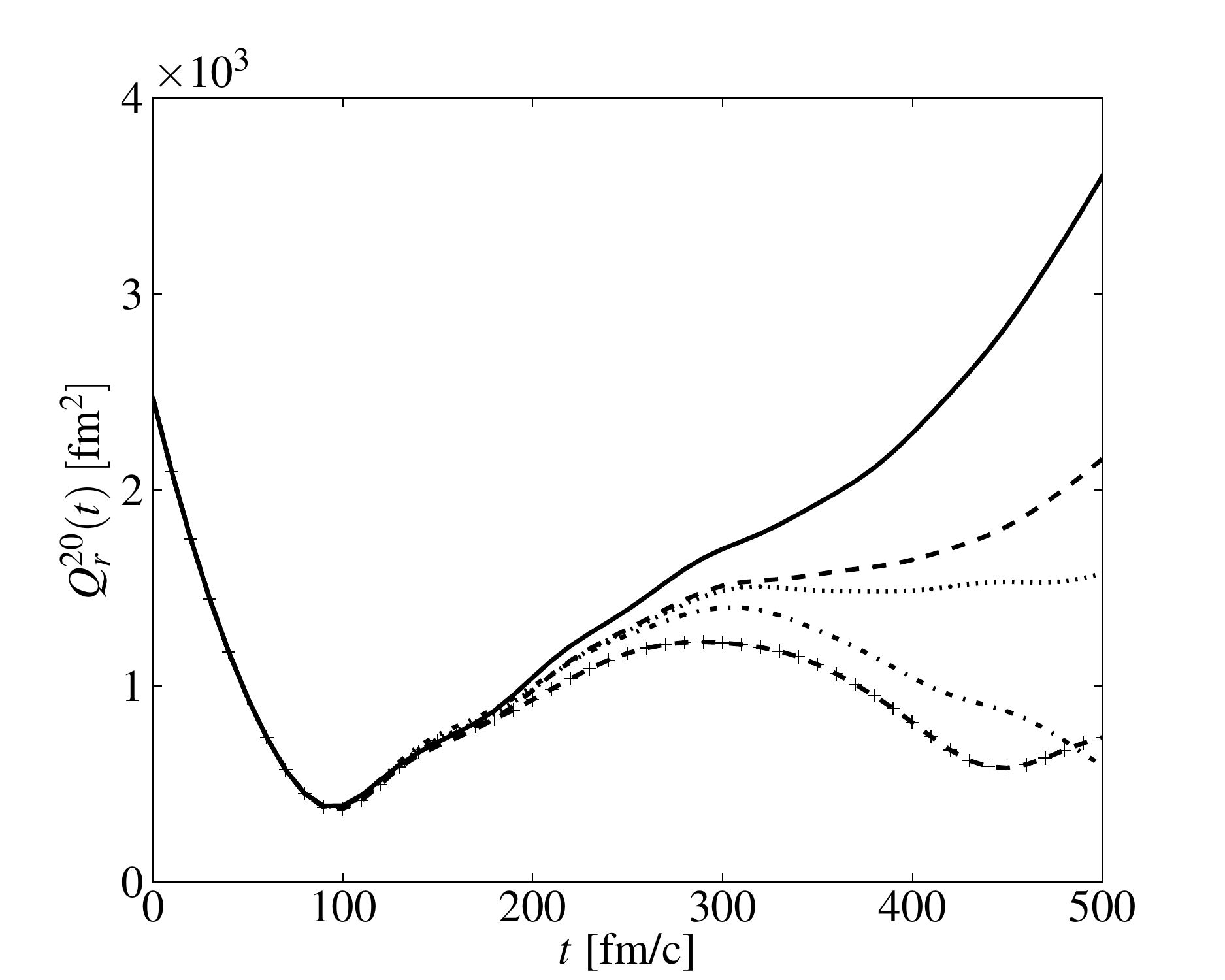}}
 \begin{pspicture}(\wd\IBox,\ht\IBox)
    \rput[lb](-0.5,0.5){\usebox\IBox}
    \rput[lb](-0.3,6.5){(d)}  
  \end{pspicture}
\caption{\label{fig:Skyrme}
Global observables $\beta_k(t)$ (a), $\gamma_k(t)$ (b), $V_k(t)$ (c),
and $Q^{20}_r(t)$ (d) are shown for a central $^{40}$Ca+$^{40}$Ca
collision with a center-of-mass energy $E_\mathrm{c.m.}=160\;$MeV. Each curve
corresponds to a different Skyrme force as indicated in the legend. 
}
\end{figure}
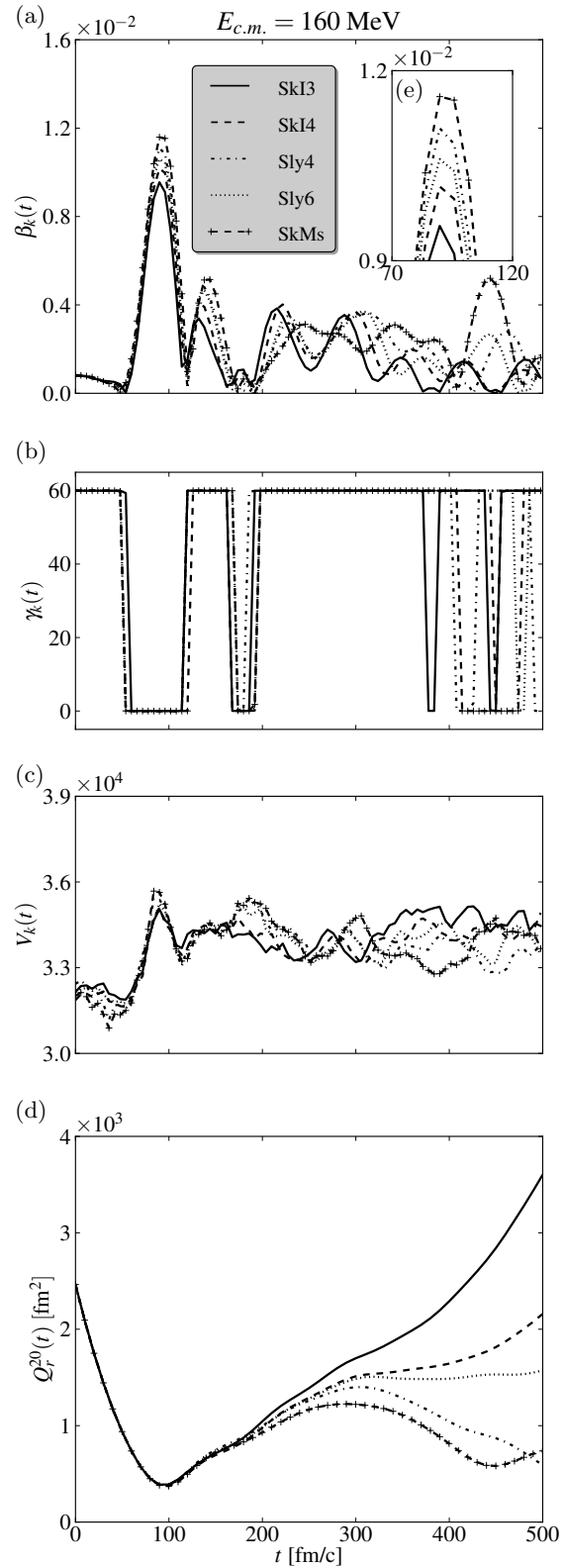

\subsection{$^{40}$Ca+$^{40}$Ca}
\label{sec:caca}

We now investigate the relaxation process in momentum space
for heavy-ion reactions. We have chosen the
$^{40}$Ca+$^{40}$Ca-system as a benchmark to investigate the
quantitative effect of some parameters on the relaxation process. All
calculations in this section were done for central collisions
(impact parameter $b=0$). The numerical grid was set up with
$36\times24^2$ grid points.

\subsubsection{Variation of the Skyrme force}

In the first set of calculations we vary the Skyrme parametrization.
Figure~\ref{fig:Skyrme} shows the results of a central
$^{40}$Ca+$^{40}$Ca collision with the Skyrme parametrizations SLy4,
SLy6 \cite{Chabanat}, SkMs \cite{Bartel}, SkI3, and SkI4 \cite{Reinhard2}.
While SkMs was chosen as an example for an outdated
interaction, the SLy(X) set of forces was originally
developed to study isotopic trends in neutron rich nuclei and neutron
matter with applications in astrophysics. The SkI(X) forces take the
freedom of an isovector spin-orbit force into account. This results in
an improved description of isotopic shifts of r.m.s. radii in
neutron-rich Pb isotopes.

The global development of the reaction is visualized in subplot (d).
The time-dependent expectation value $Q^{20}_r(t)$ shows the five
trajectories initially in good agreement but finally fanning out. A
similar splitting behavior 
depending on the employed Skyrme parametrization was already found in
\cite{Maruhn85}. While the two Sk(X)-forces
show a full separation of the two fragments, there is a slight
remaining contact between the fragments for the case of SLy6, which
will result in complete
separation in a longer calculation. However the trajectories for SLy4
and SkMs show a merged system in the final state, which was found to
persist in long-time simulations.

We now consider the relationship between the observed characteristics in coordinate space
with the dynamics in phase space. Subplot (a) shows the
$\beta_k$-value, measuring the global deviation of the momentum
distribution from a sphere. The initial
$\beta_k$-peak is strongly damped for all five Skyrme-forces. While
the time development for all parametrizations
remain in phase up to the second peak, later it starts to vary and 
continue with damped oscillations.
For a better visualization the first peak is magnified in subplot (e). The taller
the $\beta_k$-peak the longer the fragments will stick together in coordinate space. The 
effect appears to depend on the effective mass $m*/m$. Smaller effective
masses give rise to a smaller $\beta_k$-deformation. Table \ref{tab:effectm}
summarizes the $m*/m$-values associated with the
maximal deformations $\beta_k^{max}$ for all the
Skyrme forces used in this work.
\begin{table}
\caption{\label{tab:effectm}
Effective masses of Skyrme parametrizations used in this work are
listed in connection with the maximal $\beta_k$-values from the plots
(a, e) in Fig.~\ref{fig:Skyrme}.}
\begin{ruledtabular}
\begin{tabular}{lll}
Skyrme force &$m*/m$& $\beta_k^{max}$\\ 
\hline
SkM∗ &0.79&0.0116 \\
SLy4 &0.70&0.0111 \\
SLy6 &0.69&0.0106 \\
SkI4 &0.65&0.0102 \\
SkI3 &0.57&0.0095 \\
\end{tabular}
\end{ruledtabular}
\end{table}
Subplot (b) shows the $\gamma_k$-value which indicates, whether a
deformation is prolate, oblate, or triaxial \cite{Greiner}.
For the present scenario of a central collision the $\gamma_k$-value
jumps between prolate and oblate configurations indicating that the
momentum distribution oscillates between being aligned primarily in
the beam direction or transverse to it. For the sake of completeness
we additionally present the occupied
phase-space volume (c) which will prove more useful for the next
reaction parameter to be discussed: the center-of-mass energy.

\subsubsection{Variation with the center-of-mass energy}

\begin{figure}[hbtp]
 \centering
  \savebox\IBox{\includegraphics[width=8.1cm]{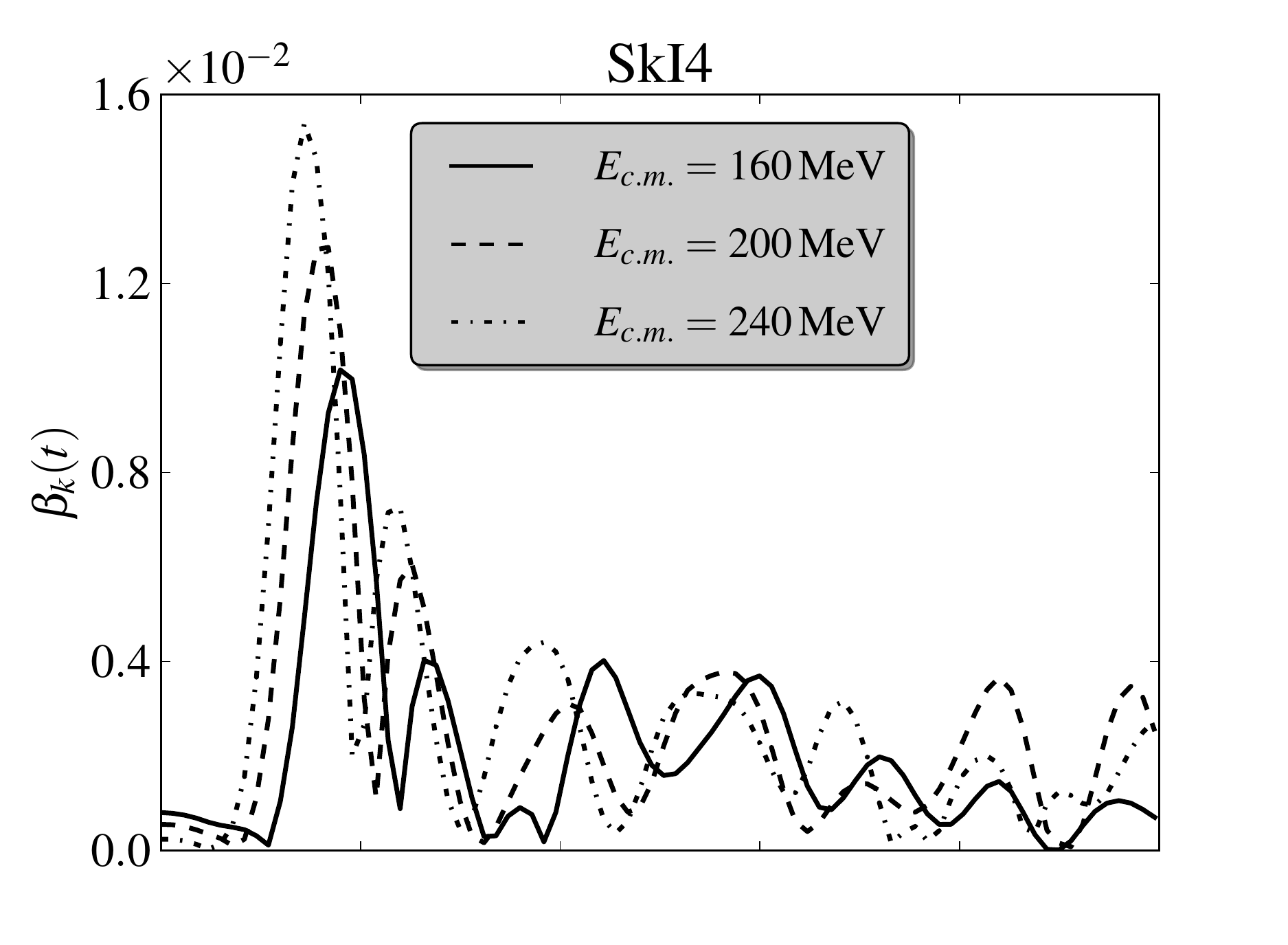}}
 \begin{pspicture}(\wd\IBox,\ht\IBox)
    \rput[lb](-0.5,0.5){\usebox\IBox}
    \rput[lb](-0.3,6.){(a)}  
  \end{pspicture}
  \savebox\IBox{\includegraphics[width=8.1cm]{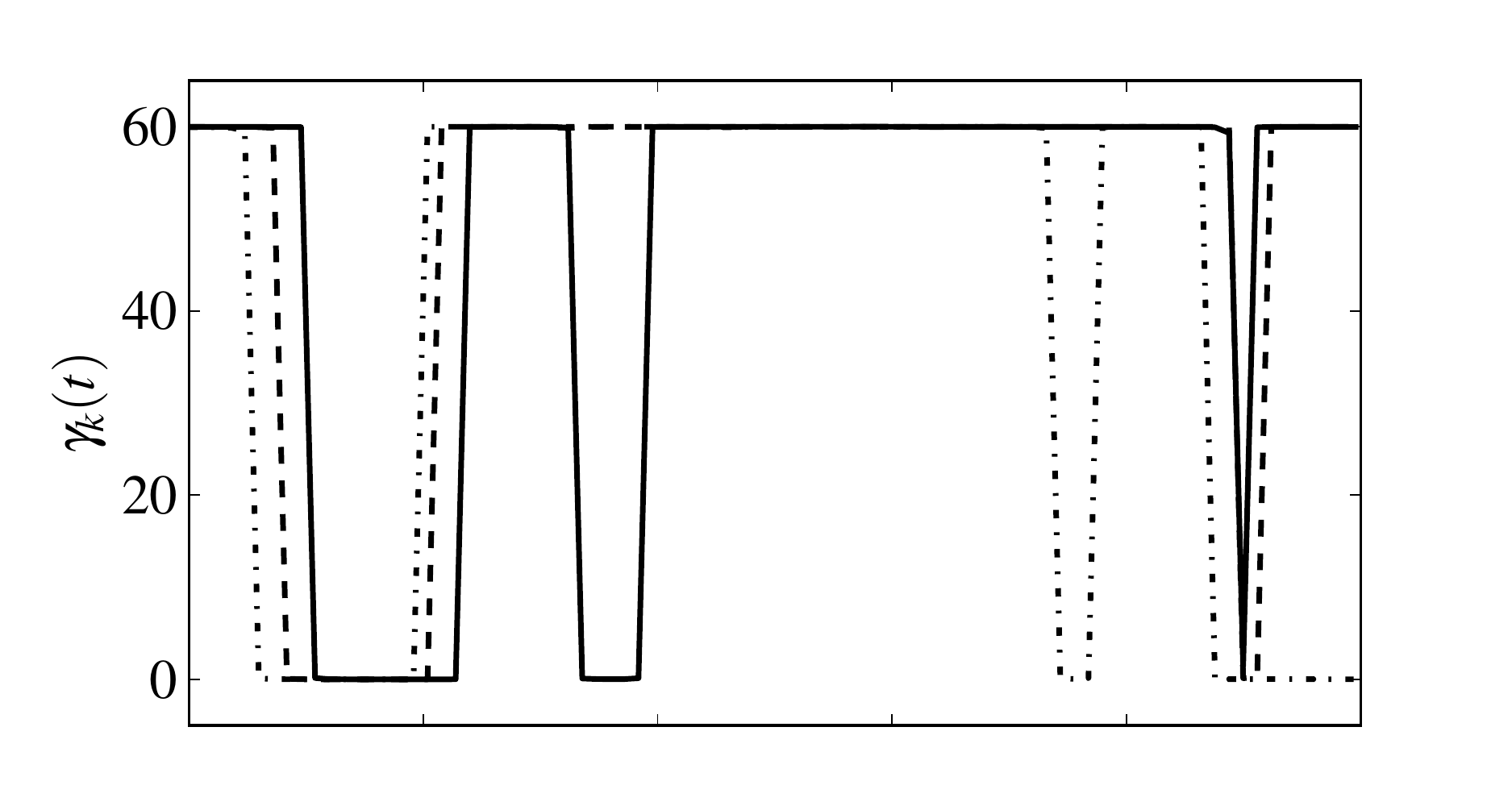}}
 \begin{pspicture}(\wd\IBox,\ht\IBox)
    \rput[lb](-0.5,0.5){\usebox\IBox}
    \rput[lb](-0.3,4.5){(b)}    
  \end{pspicture}
  \savebox\IBox{\includegraphics[width=8.1cm]{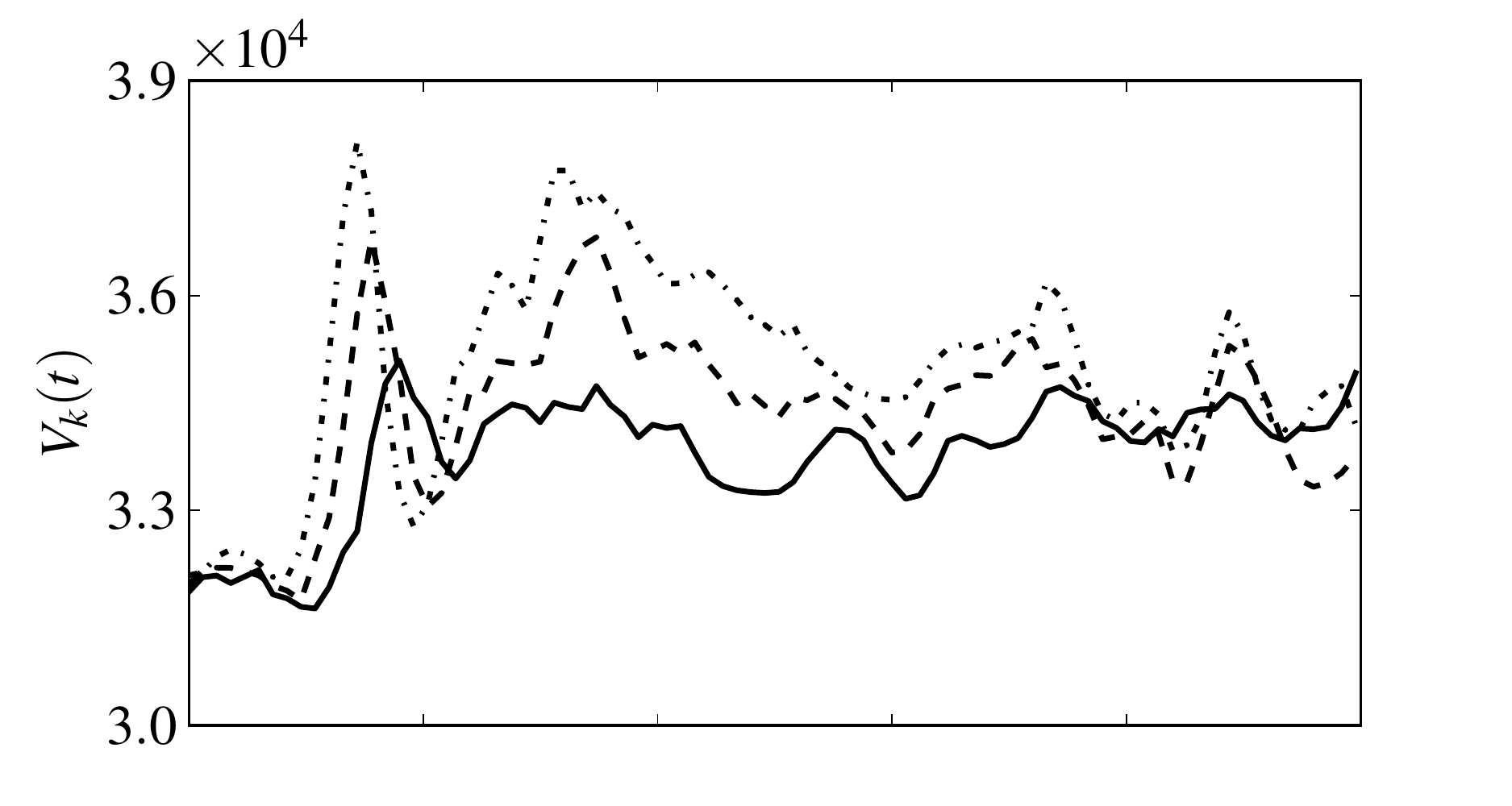}}
 \begin{pspicture}(\wd\IBox,\ht\IBox)
    \rput[lb](-0.5,0.5){\usebox\IBox}
     \rput[lb](-0.3,4.5){(c)}  
  \end{pspicture}
  \savebox\IBox{\includegraphics[width=8.1cm]{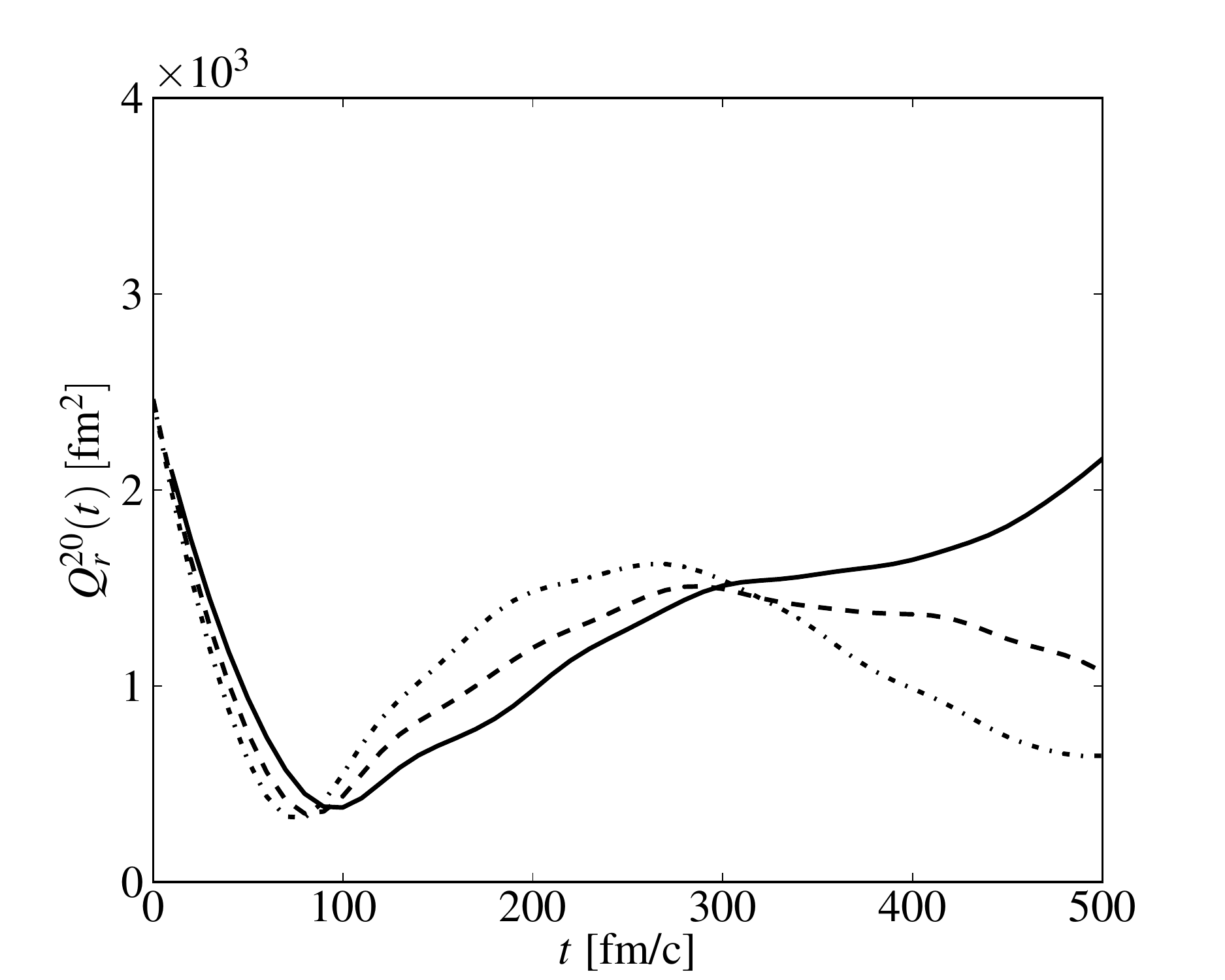}}
 \begin{pspicture}(\wd\IBox,\ht\IBox)
    \rput[lb](-0.5,0.5){\usebox\IBox}
    \rput[lb](-0.3,6.5){(d)} 
  \end{pspicture}
\caption{\label{fig:ecm} 
Global observables $\beta_k(t)$ (a), $\gamma_k(t)$ (b), $V_k(t)$ (c),
and $Q^{20}_r(t)$ (d) are shown for a central $^{40}$Ca+$^{40}$Ca
collision with fixed Skyrme interaction SkI4. Each curve corresponds
to a different center-of-mass energy.}
\end{figure}
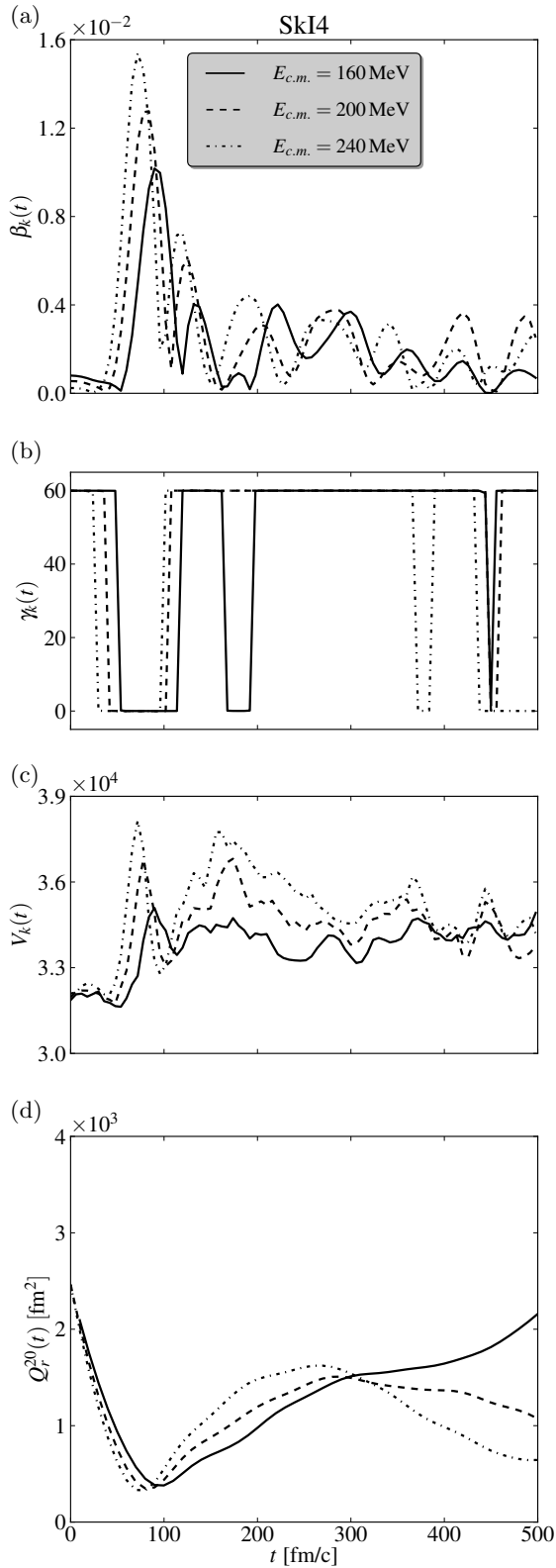

As a second reaction parameter the center-of-mass energy, $E_\mathrm{c.m.}$, is
varied. Results are presented for energies ranging from
$E_\mathrm{c.m.}=2$\:MeV/nucleon up to $E_\mathrm{c.m.}=3$\:MeV/nucleon. The Skyrme
interaction now is fixed to be SkI4. For the case of the lowest
(highest) energy Video~\ref{vid:ca160} (Video~\ref{vid:ca240})
provides a video visualizing the reaction in phase space.  The
calculation done with the lowest energy $E_\mathrm{c.m.}=160$\:MeV
shows two fully separated fragments in the exit channel. In contrast,
the case with the highest energy (as well as the one at an
intermediate energy) results in a merged system. The global
observables are plotted in Fig.~\ref{fig:ecm}.
It may not be obvious at first why the fragments should split for
lower energies and merge for higher ones. But the estimate for the
occupied phase-space volume $V_k(t)$ presented in subplot (c)
indicates that $V$ increases with energy. Therefore the fragments'
average distance in phase space is larger, while in compensation they can come
closer to each other in coordinate space. However, this behavior is also
dependent on the particular Skyrme force used and the presence of time-odd
terms, which is discussed in the next subsection.

\begin{video}
\includegraphics[width=8.1cm]{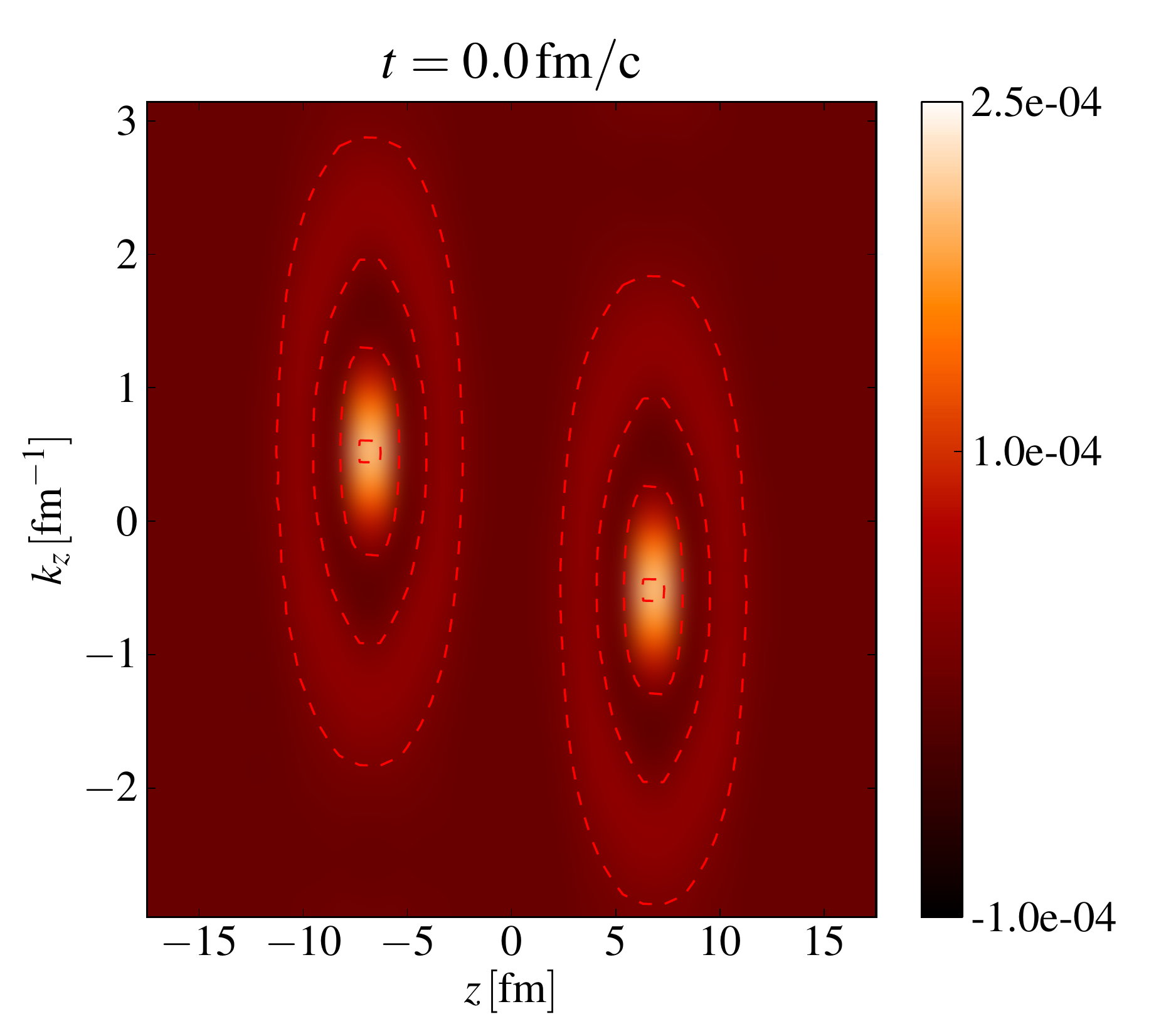}
\setfloatlink{http://th.physik.uni-frankfurt.de/~loebl/vid1.mpeg}
\caption{\label{vid:ca160} (color online)
Two-dimensional $z$-$k_z$-slice from the full six-dimensional Wigner distribution
$f^{(3)}_\mathrm{W}(\mathbf{r},\mathbf{k})$ for a central $^{40}$Ca+$^{40}$Ca collision 
with a center-of-mass energy of $E_\mathrm{c.m.}=160$\:MeV.}
\end{video}
\begin{video}
\includegraphics[width=8.1cm]{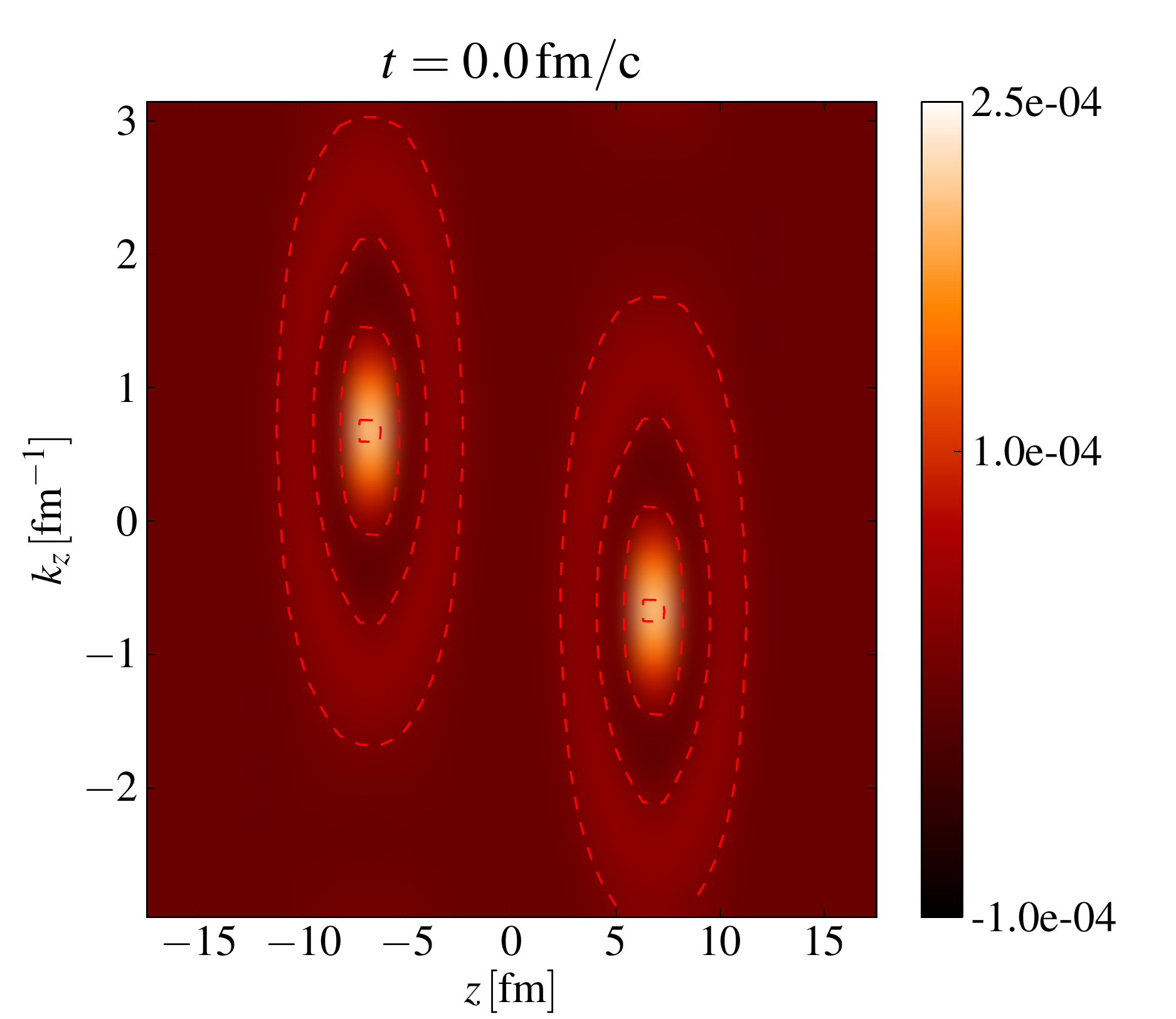}
\setfloatlink{http://th.physik.uni-frankfurt.de/~loebl/vid2.mpeg}
\caption{\label{vid:ca240} (color online)
Same as Video~\ref{vid:ca240} with a center-of-mass energy of
$E_\mathrm{c.m.}=240$\:MeV.}
\end{video}

\subsubsection{Influence of time-odd terms}

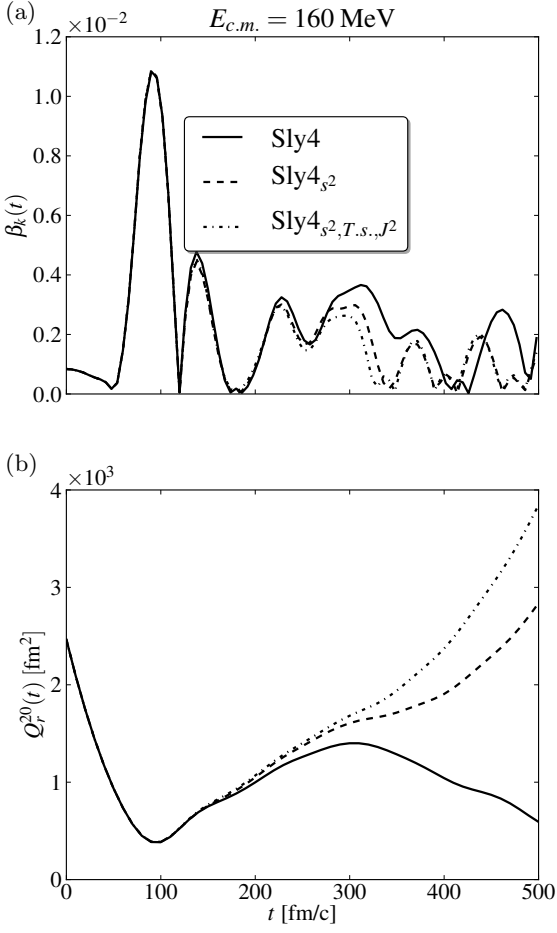
\begin{figure}[hbtp]
 \centering
  \savebox\IBox{\includegraphics[width=8.1cm]{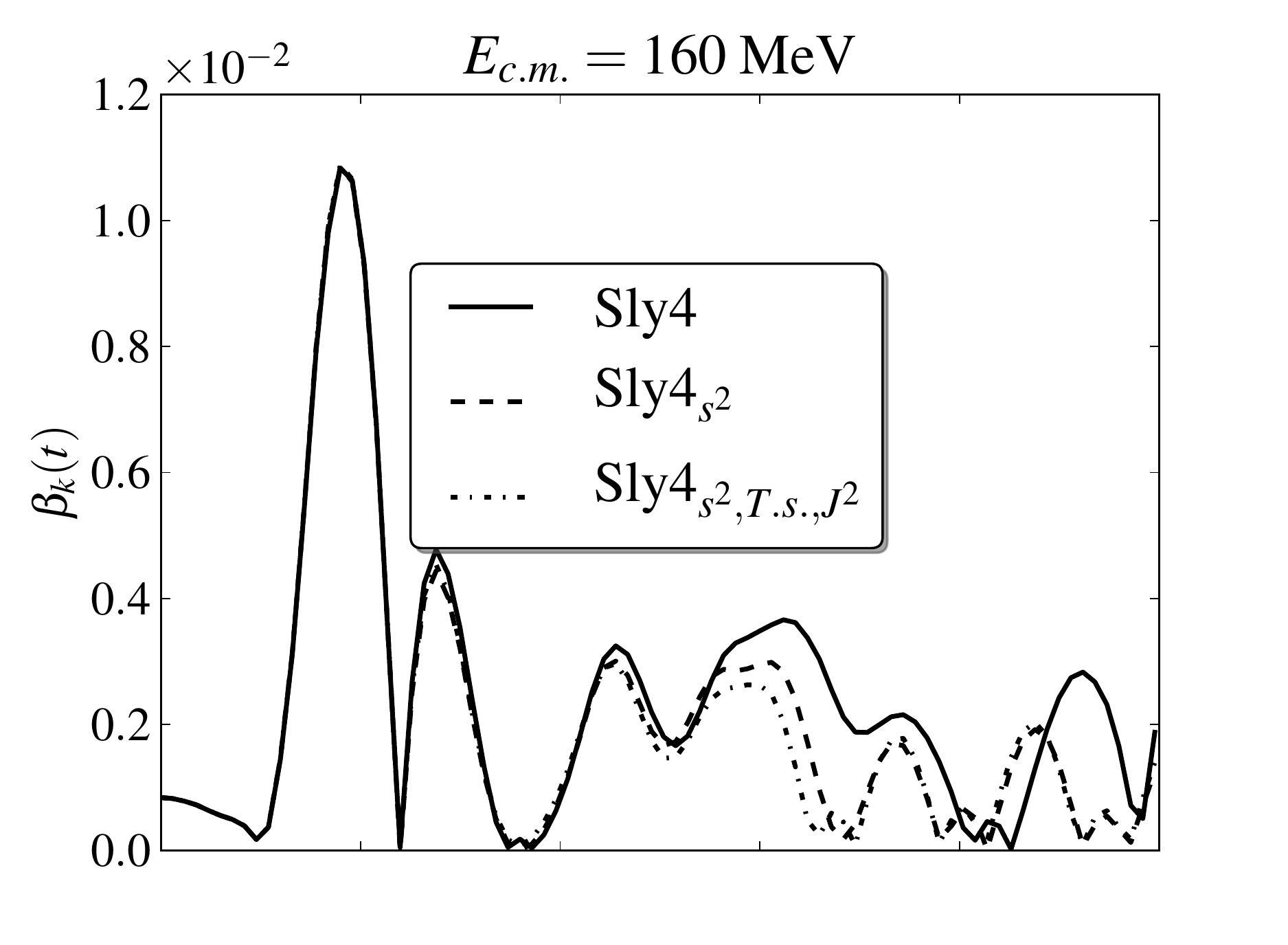}}
 \begin{pspicture}(\wd\IBox,\ht\IBox)
    \rput[lb](-0.5,0.5){\usebox\IBox}
    \rput[lb](-0.3,6.){(a)}  
  \end{pspicture}
  \savebox\IBox{\includegraphics[width=8.1cm]{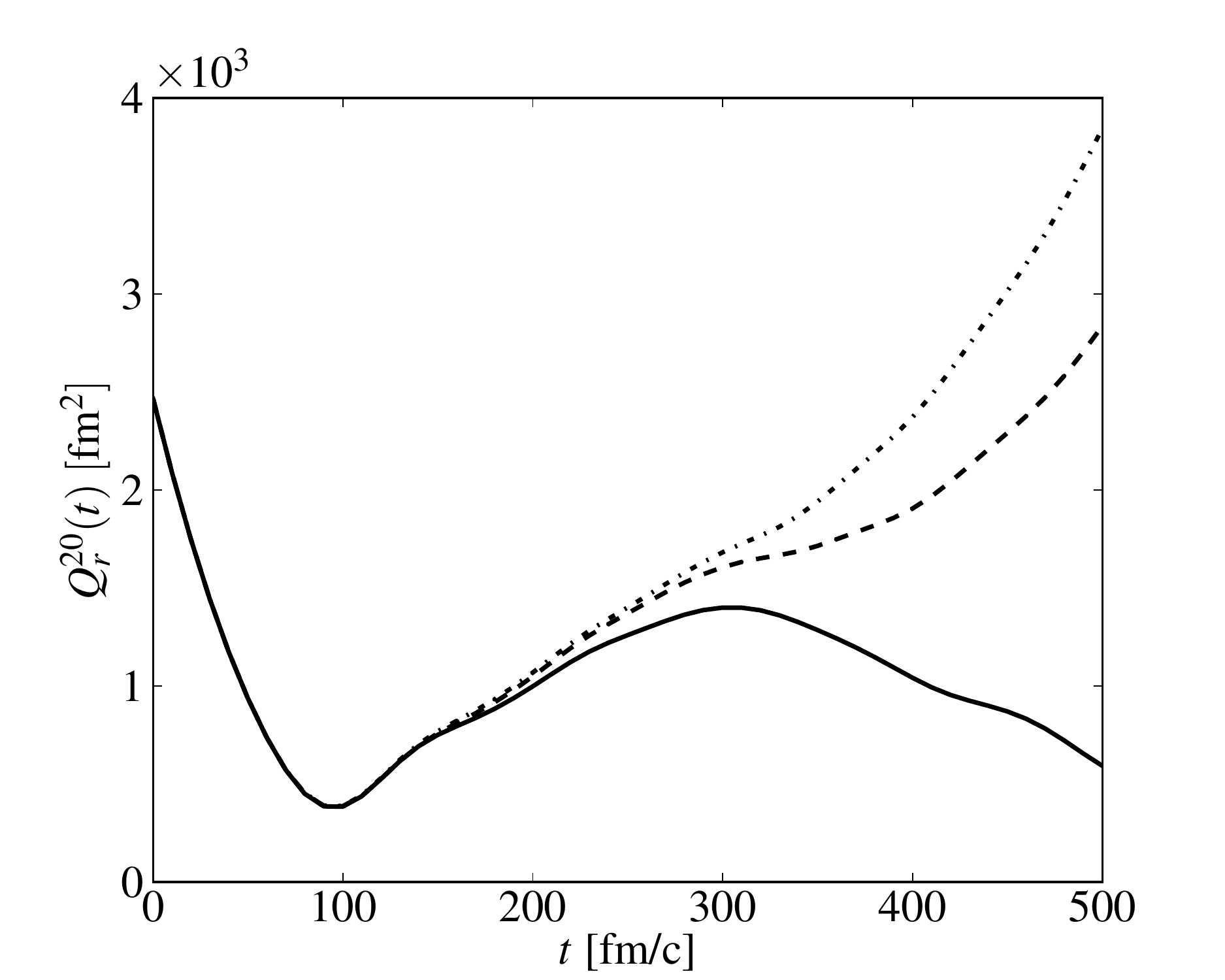}}
 \begin{pspicture}(\wd\IBox,\ht\IBox)
    \rput[lb](-0.5,0.5){\usebox\IBox}
    \rput[lb](-0.3,6.5){(b)} 
  \end{pspicture}
\caption{\label{fig:extra} 
Global observables $\beta_k(t)$ (a), and $Q^{20}_r(t)$ (d) are shown for a central $^{40}$Ca+$^{40}$Ca
collision with fixed Skyrme interaction SLy4.}
\end{figure}

Skyrme energy-density functionals are calibrated to ground state
properties of even-even nuclei~\cite{Chabanat,Bartel,Reinhard2}.
This leaves the choice of the time-odd terms in the functional (current
$\mathbf{j}^2$, spin-density $\mathbf{s}^2$, spin kinetic energy density $\mathbf{T}$,
and the spin-current pseudotensor $\tensor{\mathbf{J}}$) largely unspecified~\cite{BH03}.
Galilean invariance requires at least some of these terms to be present
depending on the presence of the associated time-even term, e.g. $\mathbf{j}^2$
for the $(\rho\tau-\mathbf{j}^2)$ combination.
In our calculations we always include the time-odd part of the spin-orbit interaction.
In order to investigate the effects of the remaining time-odd
terms, we have compared different choices by using a single Skyrme parametrization and the same test case.
We choose the force SLy4 and start with
the minimum number of time-odd terms which is needed to ensure
Galilean invariance~\cite{Eng75a}. In the next stage, we include also
the spin-density terms proportional to $\mathbf{s}^2$. Finally, we also add the combination 
which includes the tensor
spin-current term $(\mathbf{s}\cdot\mathbf{T}-\tensor{\mathbf{J}}^2)$.  As shown in Fig.~\ref{fig:extra},
at least for
the quantities $\beta_k$ and $Q^{20}_r$, varying these time-odd terms
has a very little effect in the initial contact phase and the dynamical
behavior becomes somewhat different only in later stages of the collision.
On the other hand, small differences near the threshold energy
(the highest collision energy for a head-on collision that results in a composite system.
At higher energies the nuclei go through each other)
can have large long-term effects on the outcome of the collision. 
For example a small difference in dissipation may be enough to
influence the decision between re-seperation or forming a composite system.
We have also checked a broader range of collision energies
from the fusion regime up to deep inelastic collisions. 
The interesting quantity is the loss of fragment kinetic energy
between the entrance and exit channels. It was found that the spin terms contribute
small changes to this loss which can go in both directions, less
dissipation near fusion threshold and more dissipation above.
Subplot (b) of Fig.~\ref{fig:extra} shows the effects near the Coulomb
barrier where spin terms reduce dissipation.

\subsection{$^{24}$Mg+$^{208}$Pb}
\label{sec:mgpb}

\begin{figure}[hbtp]
 \centering
 \includegraphics[width=8.1cm]{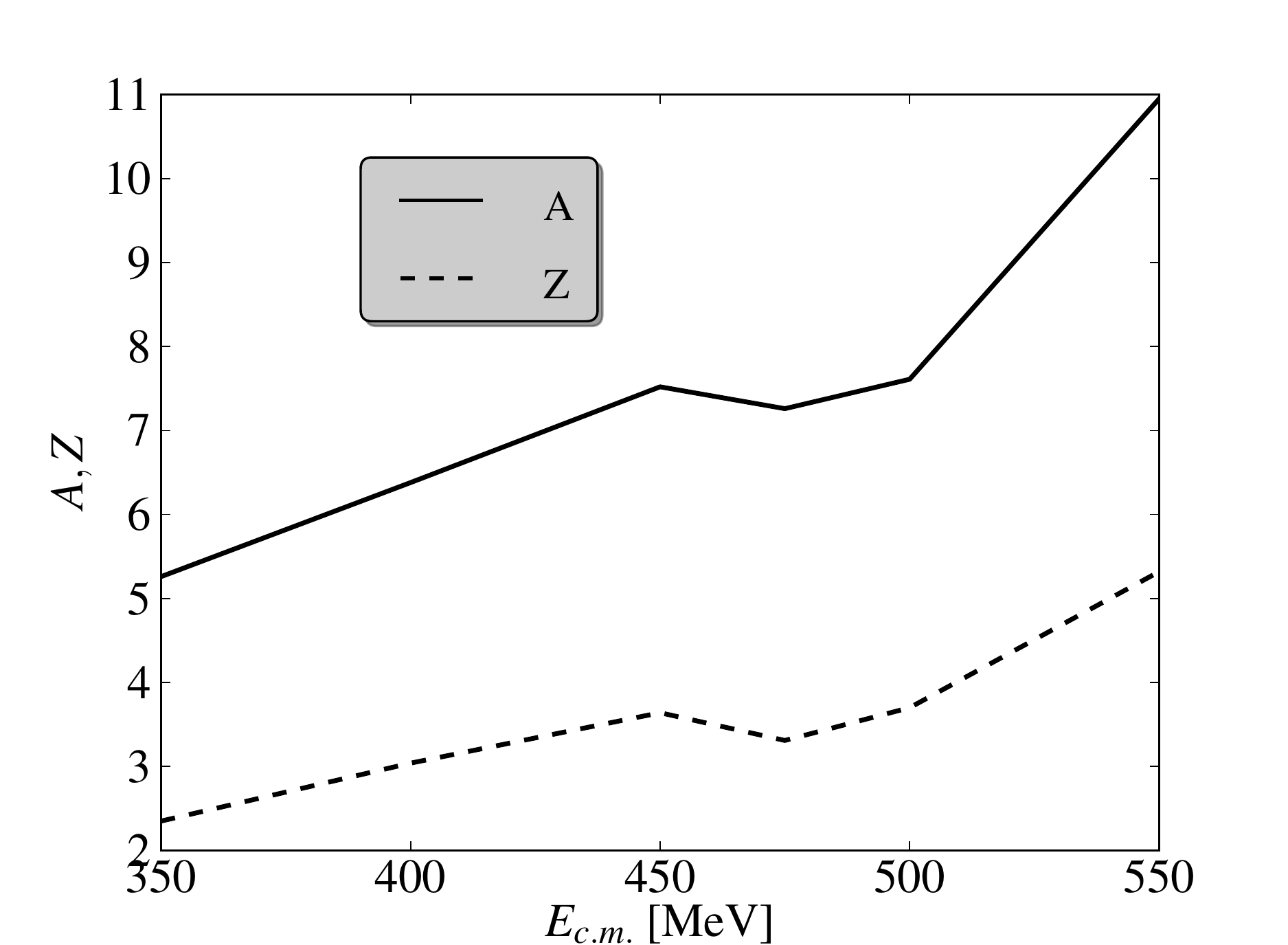}
 \caption{\label{fig:central} (color online)
Mass number $A$ and charge $Z$ of the cluster emitted in the forward direction (feed-through) for central $^{24}$Mg+$^{208}$Pb reactions 
as a function of the center-of-mass energy $E_\mathrm{c.m.}$.}
\end{figure}

The study of the $^{24}$Mg+$^{208}$Pb reaction is motivated by
experiments indicating some evaporation of heavy nucleon clusters 
around zero degree relative to the beam direction in the
$^{25}$Mg+$^{206}$Pb collision~\cite{Heinz}. The following results 
are obtained on a $32^2\times48$ grid with Skyrme interaction SLy6.

In the following we define as "feed-through" the mass $A$ and charge $Z$ numbers of the 
nuclear matter cluster emitted in the forward direction. Fig.~\ref{fig:central} shows TDHF results on the feed-through for central
$^{24}$Mg+$^{208}$Pb reactions. However, the density of the emerging fragments is very low. We
identify the energy threshold for a possible feed-through at
$E_\mathrm{c.m.}\approx 350\;$MeV.

\begin{video}
\includegraphics[width=8.1cm]{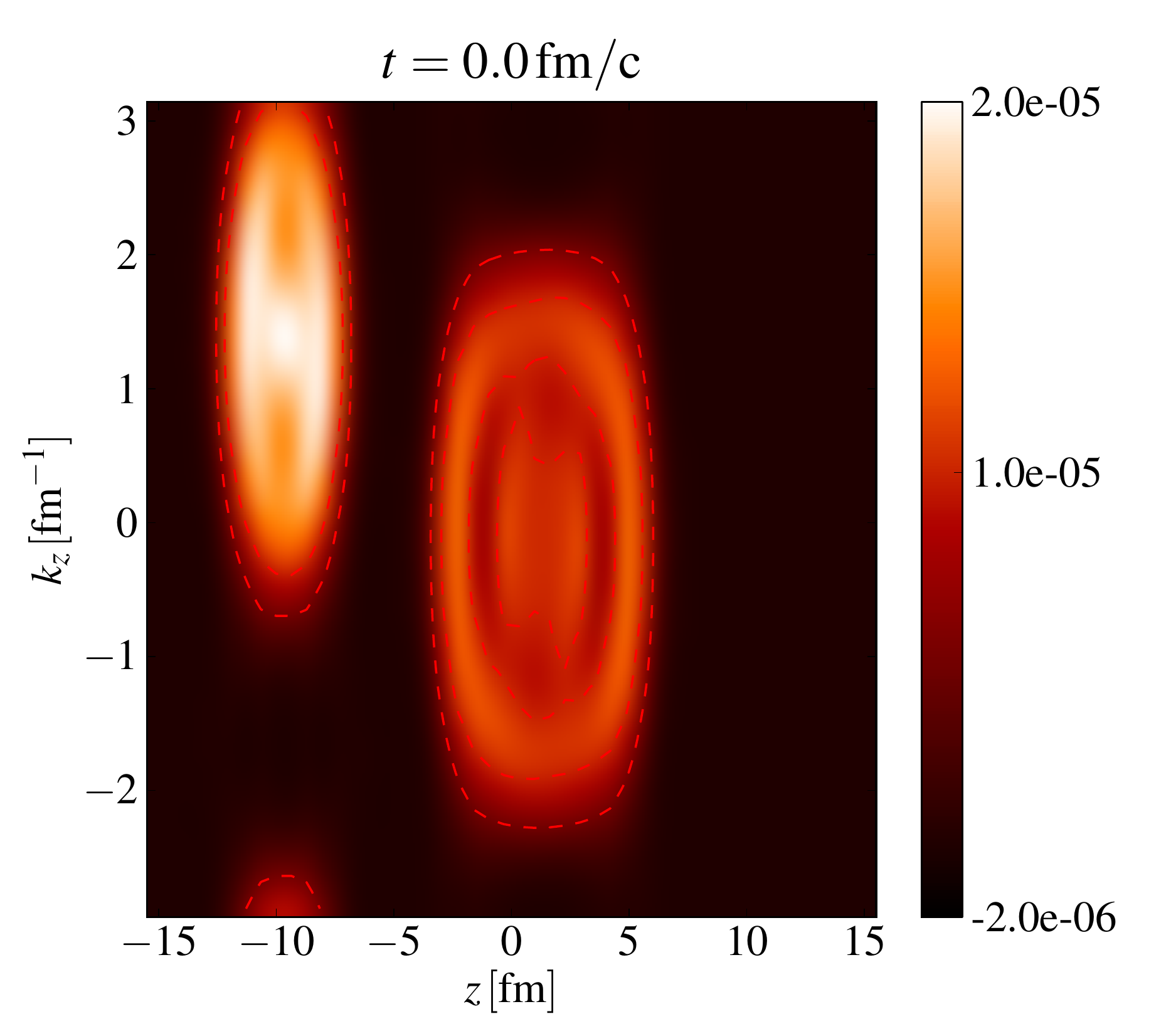}
\setfloatlink{http://th.physik.uni-frankfurt.de/~loebl/vid3.mpeg}
\caption{\label{vid:MgPbwig} (color online)
Two-dimensional $z$-$k_z$-slice from the full six-dimensional Wigner distribution
$f^{(3)}_\mathrm{W}(\mathbf{r},\mathbf{k})$ for a central $^{24}$Mg+$^{208}$Pb collision 
with a center-of-mass energy of $E_\mathrm{c.m.}=350$\:MeV.}
\end{video}

Video~\ref{vid:MgPbwig} provides a video showing the central
$^{24}$Mg+$^{208}$Pb reaction slightly above the feed-through
threshold in the Wigner picture. As time
elapses $^{24}$Mg is absorbed into the phase-space volume of
$^{208}$Pb. We observe a rotation of phase-space density 
inside the merged system. After a short time a jet is visible,
leaving the $^{24}$Mg+$^{208}$Pb compound. This corresponds to
low-density fragment in coordinate space with a total mass
comparable to an $\alpha-$particle for the actual energy. 

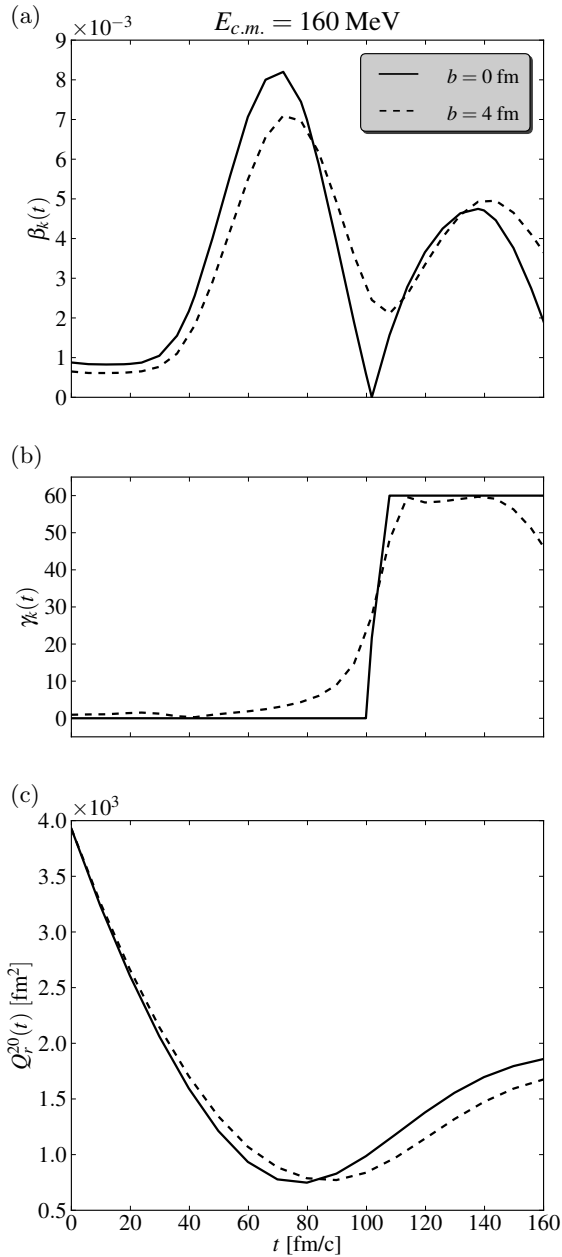
\begin{figure}[hbtp]
 \centering
  \savebox\IBox{\includegraphics[width=8.1cm]{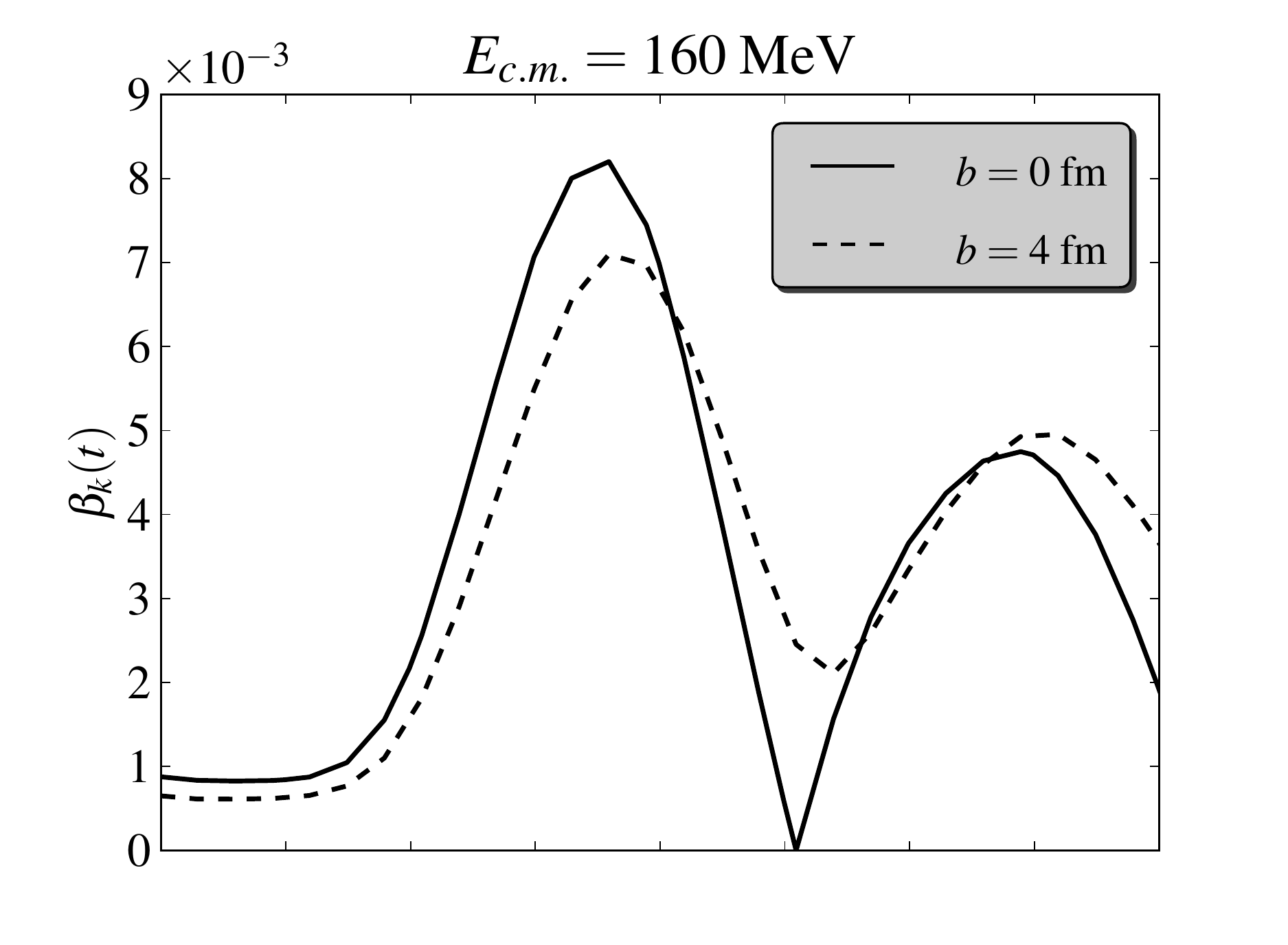}}
 \begin{pspicture}(\wd\IBox,\ht\IBox)
    \rput[lb](-0.5,0.5){\usebox\IBox}
    \rput[lb](-0.3,6.){(a)}  
  \end{pspicture}
  \savebox\IBox{\includegraphics[width=8.1cm]{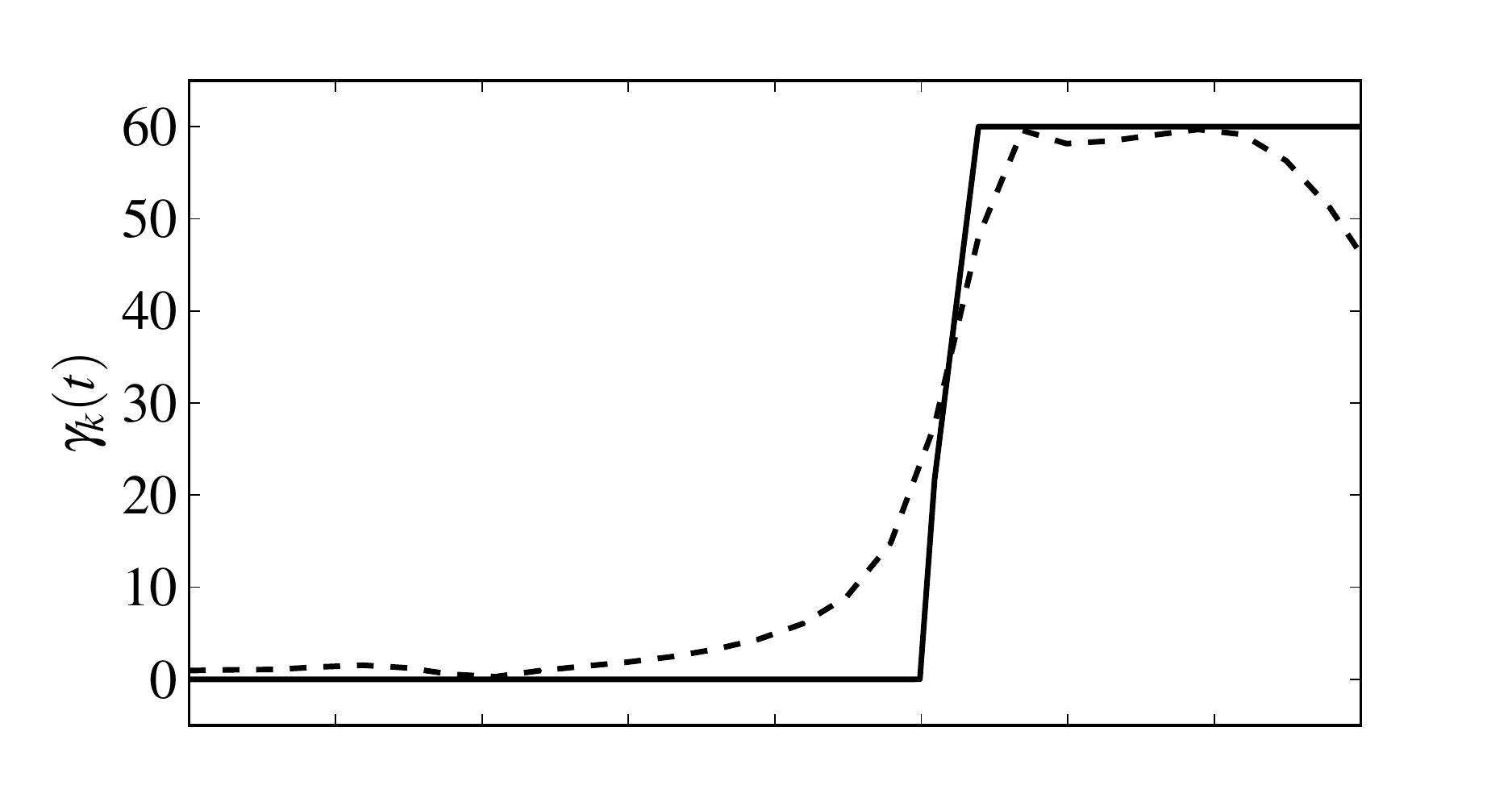}}
 \begin{pspicture}(\wd\IBox,\ht\IBox)
    \rput[lb](-0.5,0.5){\usebox\IBox}
    \rput[lb](-0.3,4.5){(b)}  
  \end{pspicture}
  \savebox\IBox{\includegraphics[width=8.1cm]{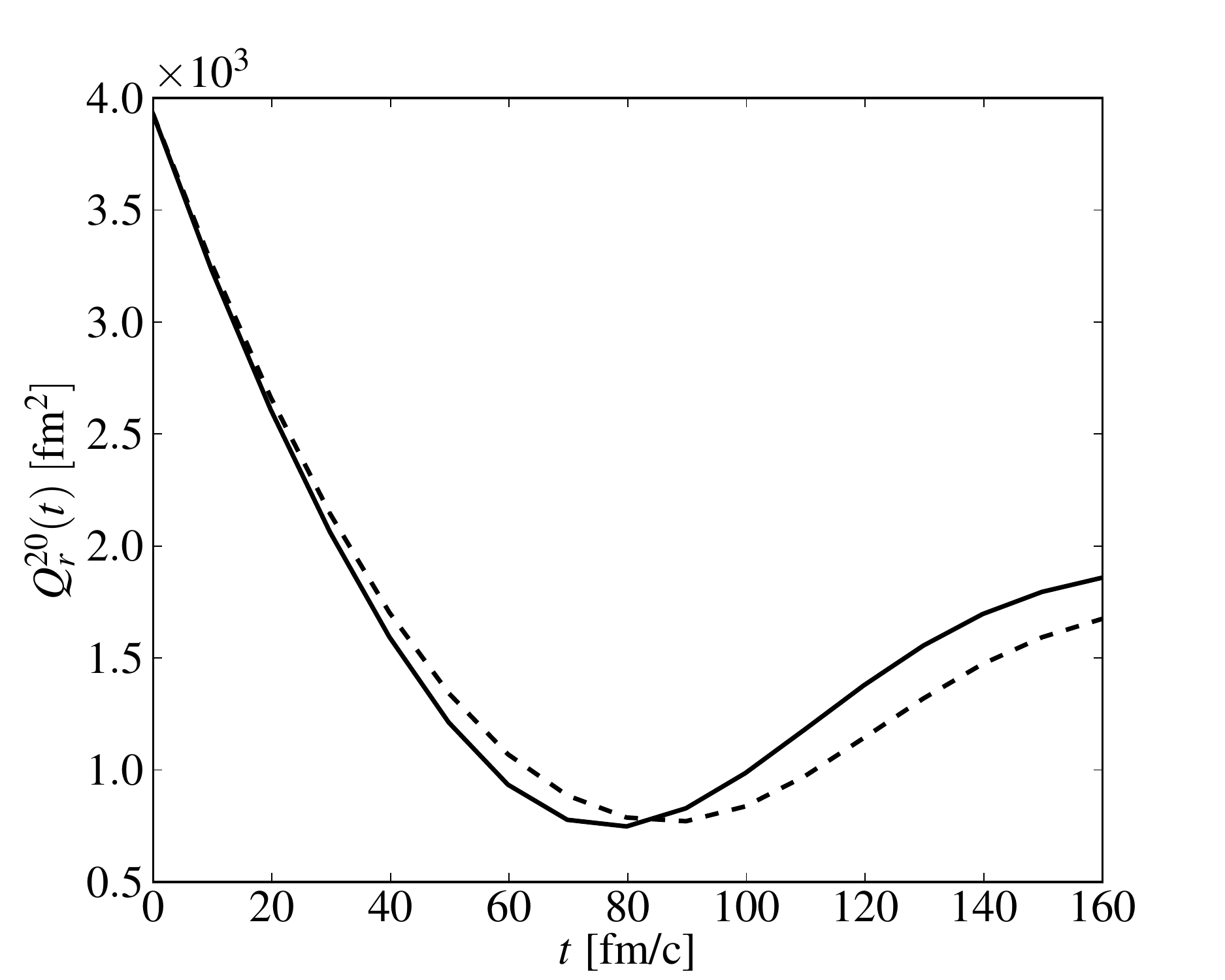}}
 \begin{pspicture}(\wd\IBox,\ht\IBox)
    \rput[lb](-0.5,0.5){\usebox\IBox}
    \rput[lb](-0.3,6.5){(c)} 
  \end{pspicture}
\caption{\label{fig:MgPb} 
Global observables $\beta_k(t)$ (a), $\gamma_k(t)$ (b), $V_k(t)$ (c),
and $Q^{20}_r(t)$ (c) are shown for a  $^{24}$Mg+$^{208}$Pb
collision. The full curve corresponds to a central reaction and the dashed
curve to a non-central reaction with impact parameter $b=4$\:fm. Both
cases are calculated with Skyrme force SLy6 and with a center-of-mass
energy $E_\mathrm{c.m.}=350$\;MeV.}
\end{figure}

In Fig.~\ref{fig:MgPb} we compare the global observables for a central
and a peripheral reaction with impact parameter $b=4$\:fm. Subplot (a)
shows that the $\beta_k$-peaks are diminished for the case of the
non-central collision. In this case we find that the alignment of the
momentum distribution probed via the $\gamma_k$-angle (b) no longer
shows a sharp transition between a prolate and an oblate
configuration. The smoother development of $\gamma_k$ is due to the
possibility for triaxial configurations in the non-central collision.

\begin{video}
\includegraphics[width=8.1cm]{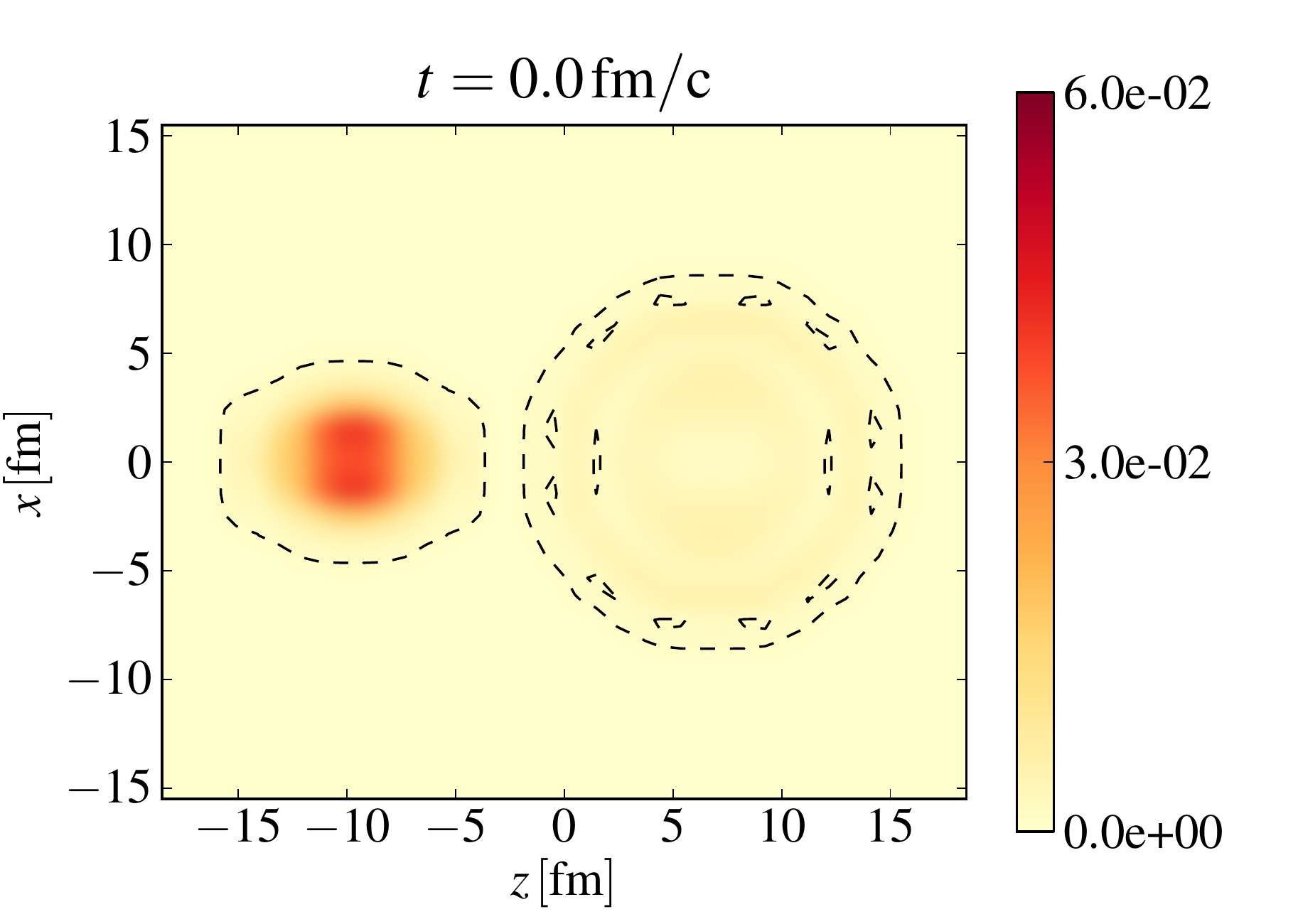}
\setfloatlink{http://th.physik.uni-frankfurt.de/~loebl/vid4.mpeg}
\caption{\label{vid:beta} (color online)
The local observable $\beta_k(\mathbf{r},t)$ is plotted in the
reaction plane,
i.e. $\mathbf{r}=(x,y=0,z)$ for a  $^{24}$Mg+$^{208}$Pb collision. The
calculation is done
with Skyrme force SLy6 and a center-of-mass energy of
$E_\mathrm{c.m.}=350$\;MeV.}
\end{video}

The time development of the local observable $\beta_k(\mathbf{r})$
can be studied in Video~\ref{vid:beta}. At $t=0$\;fm/c we observe a
non vanishing local $\beta_k$ deformation in the $^{24}$Mg fragment.
As we have previously shown in Fig.~\ref{fig:Static} (b), $^{24}$Mg 
reveals a ground state deformation in coordinate as well as in
momentum-space. In
the reaction plane a domain of high $\beta_k$ values occurs while
$^{24}$Mg is penetrating 
through the $^{208}$Pb fragment. The initial disturbance travels
through
the lead fragment while it exhibits a strong decrease in magnitude.
When the low density nuclear
cluster leaves the merged system, the $\beta_k$ excitation has
completely subsided.

\section{Summary}
\label{sec:conclusion}

We have presented a geometrically unrestricted framework to study
nuclear dynamics within TDHF in the full six-dimensional phase space.
The impact of different reaction parameters on the outcome of a
heavy-ion collision was studied in detail for $^{40}$Ca+$^{40}$Ca 
and $^{24}$Mg+$^{208}$Pb systems.
We find that there is transparency in both collisions, which is
clearly reflected in the global asymmetry of the Wigner momentum
distribution.
The surprising result that in some cases the system merges at higher
energies and shows transparency at lower ones can be related to the 
interplay between momentum- and configuration-space volumes which is a
reflection of the Pauli principle.
It is also interesting that the two distributions in phase-space never
truly combine to form a single distribution. 
This clearly indicates that two-body collisions will be necessary to
achieve true equilibrium as the reaction proceeds to longer contact
times.
The detailed degree of relaxation found depends on energy and also the
properties of the Skyrme force, 
where especially the effective mass seems to be important. The
presence of additional time-odd terms in the Skyrme functional appears
to have
a complex impact on the outcome of a collision as well. 
In this paper only one non-central collision was studied. 
A systematic investigation of impact parameter and energy dependence
as well as even heavier systems would be highly 
interesting but is beyond computational feasibility at the moment.

\section*{Acknowledgment}

This work has been supported by  by BMBF under
contract Nos. 06FY9086 and 06ER142D, and the U.S. Department
of Energy under grant No. DE-FG02-96ER40963 with
Vanderbilt University. The videos linked in the manusscript can be
found at \url{http://th.physik.uni-frankfurt.de/~loebl/vid1.mpeg},
\url{http://th.physik.uni-frankfurt.de/~loebl/vid2.mpeg},
\url{http://th.physik.uni-frankfurt.de/~loebl/vid3.mpeg}, and
\url{http://th.physik.uni-frankfurt.de/~loebl/vid4.mpeg}.

\appendix
\section{The full Skyrme functional}

Following the convention used in Ref.~\cite{Les07} we can write down
the full Skyrme energy density functional
$\mathcal{E}_{\text{Skyrme}}$. It depends on seven local densities and
currents, namely the spatial density $\rho_q (\mathbf{r})$
(time-even), the kinetic density $\tau_q (\mathbf{r})$ (time-even),
the current density $\mathbf{j}_q (\mathbf{r})$ (time-odd), the spin
density $\mathbf{s}_q (\mathbf{r})$ (time-odd), the spin-current
density $J_{q,\mu \nu}(\mathbf{r})$ (time-even), and the tensor
kinetic density $F_{q, \mu} (\mathbf{r})$ (time-odd). They are defined
as
\begin{eqnarray}
 \rho_q(\mathbf{r})&=&\rho_q(\mathbf{r},\mathbf{r}')
  \big|_{\mathbf{r}=\mathbf{r}'}\nonumber\\
 \mathbf{s}_q(\mathbf{r})&=&\mathbf{s}_q(\mathbf{r},\mathbf{r}')
   \big|_{\mathbf{r}=\mathbf{r}'}\nonumber\\
 \tau_q(\mathbf{r})&=&\boldsymbol{\mathbf{\nabla}}\cdot
   \boldsymbol{\mathbf{\nabla}}'\:\rho_q(\mathbf{r},\mathbf{r}')
   \big|_{\mathbf{r}=\mathbf{r}'}\nonumber\\
 T_{q,\mu}(\mathbf{r})&=&\boldsymbol{\mathbf{\nabla}}\cdot
  \boldsymbol{\mathbf{\nabla}}'\:s_{q,\mu}(\mathbf{r},\mathbf{r}')
  \big|_{\mathbf{r}=\mathbf{r}'}\nonumber\\
  \mathbf{j}_q(\mathbf{r})&=&-\frac{i}{2}
  (\boldsymbol{\mathbf{\nabla}}-\boldsymbol{\mathbf{\nabla}}')\:
  \rho_q(\mathbf{r},\mathbf{r}')\big|_{\mathbf{r}=\mathbf{r}'}
  \nonumber\\
 J_{q,\mu\nu}(\mathbf{r})&=&-\frac{i}{2}(\nabla_\mu-
  \nabla_\mu^\prime)\:s_{q,
  \nu}(\mathbf{r},\mathbf{r}')\big|_{\mathbf{r}=
  \mathbf{r}'}\\
 F_{q,\mu}(\mathbf{r})&=&\frac{1}{2}
  \sum_{\nu=x}^{z}
  \big(\nabla_\mu\nabla_\nu^\prime+\nabla_\mu^\prime\nabla_\nu
  \big)\:s_{q,\nu}(\mathbf{r},\mathbf{r}')\big|_{\mathbf{r}
   =\mathbf{r}'}\nonumber\:,
\end{eqnarray}
with
\begin{eqnarray}
 \rho_q(\mathbf{r},\mathbf{r}')&=&\sum_{\sigma}\rho_q(\mathbf{r}
   \sigma,\mathbf{r}'\sigma)\nonumber\\
  \mathbf{s}_q(\mathbf{r},\mathbf{r}')&=&\sum_{\sigma\sigma'}
    \rho_q(\mathbf{r}\sigma,\mathbf{r}'\sigma')\:\langle\sigma'
    |\hat{\boldsymbol{\mathbf{\sigma}}}|\sigma\rangle\:.
\end{eqnarray}
We will consider the recoupled forms of the proton and neutron
densities to isoscalar and isovector densities. The recoupling of the
density $\rho_q (\mathbf{r})$, for example, yields
\begin{eqnarray}
\rho_{0}(\mathbf{r})&=&\rho_n(\mathbf{r})+\rho_p(\mathbf{r})\:,
 \nonumber\\
\rho_{1}(\mathbf{r})&=&\rho_n(\mathbf{r})-\rho_p(\mathbf{r})\:.
\end{eqnarray}
The Skyrme full functional now reads
\begin{eqnarray}
\mathcal{E}_{\text{Skyrme}}&=&\int \D^3r\sum_{t=0,1}
\bigg\{C^\rho_t[\rho_0]\:\rho_t^2
+C^s_t[\rho_0]\:\mathbf{s}_t^2\nonumber\\
&&+C^{\Delta\rho}_t\rho_t\Delta\rho_t
+C^{\nabla s}_t(\nabla\cdot\mathbf{s}_t)^2\nonumber\\
&&+C^{\Delta s}_t\mathbf{s}_t\cdot\Delta\mathbf{s}_t
+C^\tau_t(\rho_t\tau_t-\mathbf{j}_t^2)\nonumber\\
&&+C^{T}_t\Big(\mathbf{s}_t\cdot\mathbf{T}_t
-\sum_{\mu,\nu=x}^{z}J_{t,\mu\nu}J_{t,\mu
\nu}\Big)\nonumber\\
&&+C^{F}_t\Big[\mathbf{s}_t\cdot\mathbf{F}_t
-\frac{1}{2}\Big(\sum_{\mu=x}^{z}J_{t,\mu
\mu}
\Big)^2\nonumber\\
&&-\frac{1}{2}
\sum_{\mu,\nu=x}^{z}J_{t,\mu\nu}J_{t,\nu
\mu}
\Big]
\nonumber\\
&&
+C^{\nabla\cdot J}_t(\rho_t\nabla\cdot\mathbf{J}_t
+\mathbf{s}_t\cdot\nabla\times
\mathbf{j}_t)
\bigg\}\:.
\end{eqnarray}
The functional accounts for the central, tensor, and spin-orbit
interaction. All possible bilinear terms up to second order in the
derivatives are included. The coupling constants of the central and
tensor part are listed below in terms of the well known Skyrme
parameters
\begin{eqnarray}
 C_0^{\rho}&=&\frac{3}{8}t_0+ \frac{3}{48} t_3 \:
  \rho_0^\alpha(\mathbf{r})\nonumber\\
 C_1^{\rho}&=& - \frac{1}{4}  t_0 \big( \frac{1}{2} + x_0 \big)
  -\frac{1}{24} t_3 \big( \frac{1}{2} + x_3 \big)
  \:\rho_0^\alpha (\mathbf{r})\nonumber \\
 C_0^{s}&=&- \frac{1}{4} t_0 \big( \frac{1}{2} - x_0 \big)
  -\frac{1}{24} t_3 \big( \frac{1}{2} - x_3
  \big)\:\rho_0^\alpha(\mathbf{r}) \nonumber \\
 C_1^{s}&=&-\frac{1}{8} t_0-\frac{1}{48} t_3  \:
  \rho_0^\alpha(\mathbf{r})\nonumber\\
 C_0^{\tau}&=&  \frac{3}{16}\:t_1+\frac{1}{4} t_2 \:
  \big(\frac{5}{4} + x_2 \big)\nonumber\\
 C_1^{\tau}&=&-\frac{1}{8} t_1 \big(\frac{1}{2} + x_1\big)
   +\frac{1}{8} t_2 \big(\frac{1}{2} + x_2 \big)\nonumber\\
 C_0^{T}&=&-\frac{1}{8} t_1 \big( \frac{1}{2}-x_1\big)\:
   +\frac{1}{8} t_2 \big( \frac{1}{2} + x_2 \big)
   -\frac{1}{8} (t_e + 3 t_o)    \nonumber \\
 C_1^{T}&=&-\frac{1}{16}t_1+\frac{1}{16} t_2-\frac{1}{8}
  -(t_e-t_o)\nonumber\\
 C_0^{\Delta\rho}&=&- \frac{9}{64} t_1
  +\frac{1}{16}  t_2 \big( \frac{5}{4} + x_2 \big)\nonumber\\
 C_1^{\Delta\rho}&=&\frac{3}{32}t_1 \big( \frac{1}{2} + x_1 \big)
  +\frac{1}{32} t_2 \big( \frac{1}{2} + x_2 \big)\nonumber\\
 C_0^{\Delta s}&=&  \frac{3}{32} t_1 \big( \frac{1}{2} - x_1 \big)
  +\frac{1}{32} t_2 \big( \frac{1}{2} + x_2 \big)-\frac{3}{32}
  (t_e-t_o)\nonumber\\
 C_1^{\Delta s}&=&\frac{3}{64} t_1 + \frac{1}{64}t_2 
  -\frac{1}{32}(3t_e+t_o)\nonumber\\
 B^{F}_0&=&-\frac{3}{8}(t_e+3 t_o) \nonumber\\
 B^{F}_1&=&-\frac{3}{8}(t_e-t_o) \nonumber\\
 B^{\nabla s}_0 &=&-\frac{9}{32}(t_e-t_o)\nonumber\\
 B^{\nabla s}_1 &=&-\frac{3}{32}(3t_e+t_o)\:.
\end{eqnarray}
The coupling constants of the spin-orbit part of the functional 
can be rewritten into the form
\begin{eqnarray}
 C_0^{\nabla J}&=&-\frac{3}{4} W_0\:,\nonumber \\
 C_1^{\nabla J}&=&-\frac{1}{4} W_0\:.
\end{eqnarray}


\begin{thebibliography}{00}
\bibitem{Bonche}
P. Bonche, S. E. Koonin, and J. W. Negele, Phys. Rev. 
C {\bf 13}, 1226 (1976). 

\bibitem{Svenne}
J. P. Svenne, Adv. Nucl. Phys. {\bf 11}, 179 (1979).

\bibitem{Negele}
J. W. Negele, Rev. Mod. Phys. {\bf 54}, 913 (1982).

\bibitem{Davies}
K. T. R. Davies, K. R. S. Devi, S. E. Koonin, and M. R.
Strayer, in Treatise on Heavy-Ion Physics, Vol. 3 Compound System
Phenomena, edited by D. A. Bromley (Plenum Press, New York,
1985), p. 3.

\bibitem{Kim}
K.-H. Kim, T. Otsuka, and P. Bonche, J. Phys. G {\bf 23},
1267 (1997).

\bibitem{Simenel}
C. Simenel and P. Chomaz, Phys. Rev. C {\bf 68}, 024302 (2003).

\bibitem{Nakatsukasa} T. Nakatsukasa and K. Yabana, Phys.
Rev. C {\bf 71}, 024301 (2005).

\bibitem{Umar05a}
A.~S. Umar and V.~E. Oberacker, Phys. Rev. C {\bf 71},  034314 
(2005).

\bibitem{Maruhn1} J. A. Maruhn, P.-G. Reinhard, P. D.
Stevenson, J. R. Stone, and
M. R. Strayer, Phys. Rev. C {\bf 71}, 064328 (2005).

\bibitem{Guo08a}
Lu Guo, P.--G. Reinhard, and J. A. Maruhn,
Phys. Rev. C,
{\bf 77}, 041301 (2008).

\bibitem{Wigner}
E. P. Wigner, Phys. Rev {\bf 40}, 749 (1932).

\bibitem{Loebl} N. Loebl, J. A. Maruhn, and P.-G. Reinhard, Phys. Rev.
C {\bf 84},  (2011).

\bibitem{Maruhn3}
J. A. Maruhn, Proc. Topical Conf. on Heavy-ion
collisions, Oak Ridge National Laboratory report CONF-770602, Fall
Creek Falls State Park, TN (1977).

\bibitem{Blum}
V. Blum, G. Lauritsch, J. A. Maruhn, and
P.-G. Reinhard, J. Comput. Phys. {\bf 100}, 364 (1992).

\bibitem{Reinhard}
P.-G. Reinhard and R. Y. Cusson, Nucl. Phys. {\bf A378}, 418
(1982).

\bibitem{Flocard}
H.~Flocard, S.~E.~Koonin, and M.~S.~Weiss,
Phys.\ Rev.\  {\bf C17 } (1978)  1682-1699.

\bibitem{Chabanat}
E. Chabanat, E. P. Bonche, P. Haensel, J. Meyer, and
R. Schaeffer, Nucl. Phys. A {\bf 635}, 231 (1998).

\bibitem{Bartel}
J. Bartel, P. Quentin, M. Brack, C. Guet, and H.-B. H$\mathring{\mbox{a}}$kansson,
Nucl. Phys. {\bf A386}, 79 (1982).

\bibitem{Reinhard2}
P.-G. Reinhard and H. Flocard, Nucl. Phys. A {\bf 584}, 467
(1995).

\bibitem{Maruhn85}
  J.~A.~Maruhn, K.~T.~R.~Davies, M.~R.~Strayer,
  Phys.\ Rev.\  {\bf C31}, 1289-1296 (1985).
  
\bibitem{Greiner}
W. Greiner, and J.A. Maruhn, ``Nuclear models'', Springer-Verlag, Berlin, New York (1996).

\bibitem{Heinz}
S. Heinz, V. Comas, F. P. He\ss berger, S. Hofmann, D. Ackermann, H.
G. Burkhard, Z. Gan, J. Heredia, J. Khuyagbaatar and B. Kindler, {\it
et al.}, {\it The European Physical Journal A - Hadrons and Nuclei},
Springer Berlin / Heidelberg, 2008, {\bf 38}, 227-232.

\bibitem{BH03}  M. Bender, P.-H. Heenen, and P.-G. Reinhard,
                Rev. Mod. Phys. \textbf{75}, 121 (2003).

\bibitem{Eng75a}
Y.~M. Engel, D.~M. Brink, K.~Goeke, S.~J. Krieger, and D.~Vautherin,
Nucl. Phys. A {\bf 249}, 215 (1975).

\bibitem{Les07}
T. Lesinski, M. Bender, K. Bennaceur, T. Duguet, and J. Meyer,
  Phys.\ Rev.\  {\bf C76}, 014312 (2007).


\end{thebibliography}
\end{document}